\newif\ifcbtex
\cbtexfalse

\ifcbtex

\documentclass[a4paper,12pt]{article}
\usepackage{times}
\usepackage{cb2}
\usepackage{epsf}
\usepackage{graphicx}

\titleline{C.Best: Algebraic Multigrid for Disordered Systems\dots}

\else

\documentclass[preprint,aps]{revtex4}
\usepackage{graphicx}

\newcommand{\nnm}{\nonumber}
\newcommand{\setZ}{{\bf Z}}
\newcommand{\onehalf}{\mbox{$\scriptstyle 1\over 2$}}
\newcommand{\const}{\mbox{\rm const}}

\fi

\newcommand{\conj}{+}
\renewcommand{\and}{{\&} }

\begin{document}

\preprint{HLRZ2000\_21}
\title{Algebraic Multigrid for Disordered Systems\\and Lattice Gauge
  Theories}

\ifcbtex

\author{Christoph Best\\
  {\small John von Neumann Institute for Computing/DESY,}
  {\small 52425 J\"ulich, Germany}\\
  {\small c.best@computer.org}}
\date{November 30, 2000}
\maketitle
\vspace*{-1cm}

\else

\author{Christoph Best}
\email{c.best@computer.org}
\affiliation{%
  John von Neumann Institute for Computing/DESY,
  52425 J\"ulich, Germany}

\date{November 30, 2000}

\fi

\begin{abstract}
  The construction of multigrid operators for disordered linear
  lattice operators, in particular the fermion matrix in lattice gauge
  theories, by means of algebraic multigrid and block LU decomposition
  is discussed. In this formalism, the effective coarse-grid operator
  is obtained as the Schur complement of the original matrix.  An
  optimal approximation to it is found by a numerical optimization
  procedure akin to Monte Carlo renormalization, resulting in a
  generalized (gauge-path dependent) stencil that is easily evaluated
  for a given disorder field.  Applications to preconditioning and
  relaxation methods are investigated.
\end{abstract}

\ifcbtex

\vspace*{1cm}
{\small
\noindent{\bf Table of Contents}

\tableofcontents}

\clearpage

\else

\maketitle

\tableofcontents

\fi

\section{Introduction}

Most of the computer time in lattice gauge theory simulations is spent
today on inverting the Dirac matrix, which describes the dynamics of
quarks.  While this problem can be tackled efficiently by iterative
solvers such as Krylov subspace methods
\cite{heplat-9404013,heplat-9608074,gutknecht-wup-1999}, these solvers
suffer from critical slowing down in the physically interesting
regions of small quark masses. One of the limiting factors here is
that the elementary operation in iterative solvers, the matrix-vector
multiplication, affects only next neighbors and thus limits the speed
of information propagation over the lattice (though it can be improved
by ILU preconditioning \cite{heplat-9602019,heplat-0011080}).  Using
multigrid methods, one attempts to overcome this limitation by
propagating information on a hierarchy of coarser and coarser grids.

Multigrid methods \cite{hackbusch-springer,mccormick-siam} for the
Dirac matrix inversion were inspired by Mack's multigrid approach to
quantum field theory \cite{mack-1988} and have been proposed and
investigated around 1990 by groups in Boston
\nocite{juelich-1991}
\cite{prd-43-1965,brower-moriarty-rebbi-vicari,brower-edwards-rebbi-1991,
  vyas-jue-1991,prd-vyas}, Israel
\cite{plb-253-185,solomon-jue-1991,heplat-9204014,lauwers-benav-solomon-1992},
Amsterdam \cite{hulsebos-jue-1991,
  npb-331-531,plb-272-81,amsterdam-1991}, and Hamburg
\cite{kalkreuter-jue-1991,
  heplat-9304004,heplat-9310029,heplat-9409008,heplat-9408013,thesis-baeker}.
Though it was shown that multigrid methods would, in principle,
greatly reduce or eliminate critical slowing down, they have not been
able to improve the performance of the Dirac matrix inversion in
actual applications.  These classical multigrid methods are based on a
geometrical blocking of the lattice that leads to an effective
coarse-grid formulation of the original matrix by a Galerkin approach.
The choice of the blocking and interpolation kernels is crucial for
the performance of the multigrid algorithm; they must be chosen such
that the long-range and short-range dynamics decouple as much as
possible.  An optimal choice of the blocking kernels thus depends on
the gauge field and must in principle be recalculated for each gauge
configuration as the solution of a minimization problem. As this is
not feasible, various approximations have been introduced in the
literature.


\nocite{wuppertal-1999-proceedings} 

Since then, the algebraic multigrid method \cite{siam-ruge-stueben,
  reusken-wup-1999,notay-wup-1999,brandt-2000,zhang-2000} has emerged
as a proven new method in numerical mathematics which does not rely on
a geometrical decomposition of the lattice, but solely on the matrix
itself.  It starts out from a thinned lattice that contains a certain
subset of the original lattice sites retaining their original values
and constructs an effective operator on this lattice by means of an
block $LU$ decomposition.  Thus, instead of geometric proximity, the
actual matrix elements determine in the coarsening procedure.  For
lattice gauge theory, this would automatically take the dynamics of
the gauge field, which appears as a phase factor in the matrix
elements, into account. (A similar reasoning was already cited by
Edwards, Goodman, and Sokal \cite{prl-61-1333} in one of the first
papers on multigrid methods for disordered systems, namely on random
resistor networks, though it was later assumed that algebraic
multigrid would be much more costly than the geometric variant
\cite{heplat-9204014}.) Recently, algebraic multigrid was applied to
lattice gauge theory by Medeke \cite{medeke-wup-1999}, and conversely
research about renormalization-group improved and perfect operators in
lattice gauge theory has been used to obtain efficient coarse-graining
schemes for partial and stochastic differential equations
\cite{condmat-0009449}.



One advantage of algebraic multigrid over geometric methods is that it
directly gives a prescription for calculating the coarse-lattice
operator and the interpolation kernel by a submatrix inversion
problem. For this purpose, it is instructive to consider the problem
of matrix inversion not in terms of the spectral decomposition, but in
terms of the von Neumann series
\cite{kuti-neumann-series,vyas-random-walk}.  Combined with the block
$LU$ decomposition, this results in a simple computational picture of
the method as a noninteracting random walk on a hierarchically
decomposed lattice in a disordered background. The resulting algebra
of paths on this lattice is the basis for the computational
construction of multigrid operators below, where the expansion
coefficients are calculated numerically in a process similar to Monte
Carlo renormalization. In that way, physical (or rather,
computational) information about the fine-lattice operator is
harvested to approximate the coarse-lattice operator as closely as
desired. 

The lattice fermion matrix is an example of the more general problem
of disordered matrices operators which has many applications in
different fields of computational physics.  Some recent instances in
the literature are e.g.  the transport properties of light particles
in solid-state physics \cite{kehr-1996,guo-miller-2000}, or transport
in porous media, where multigrid methods have already been employed
\cite{moulton-knapek-dendy-1998,SKnapek:99b}. The numerical method
investigated here is sufficiently general to be applicable to such
problems.

Another factor that might turn out in favor of multigrid algorithms
comes from recent advances in computer technology.  Supercomputer
architecture has long been dominated by vector-oriented machines that
can quickly execute relatively simple and uniform elementary
operations on large vectors.  But because processor speed
advances much faster than memory access speed, the performance of
such operations is now severely limited by memory bandwidth.
In computer architecture, this is overcome by using
cache-oriented architectures such as in modern microprocessors, from
which e.g.~cluster computers \cite{iwcc-melbourne} are built.
These architectures are efficient for algorithms that have a high
balance, i.e.~number of operations per memory access, and make a
frequent reuse of memory locations
\cite{iwcc-melbourne-cbest,cs-0007027}, but can tolerate much more
complexity and irregularity. Multigrid algorithms usually result in
denser, but smaller matrices and might thus be favored by these
architectures.

Finally, problems other than matrix inversion have recently come into
focus for the Dirac operator. One is the calculation of its low-lying
eigenvalues, which was one of the first applications of multigrid
\cite{heplat-9408013} and is today used for investigating hadron
structure \cite{heplat-9709130} and chiral properties
\cite{heplat-0010049,heplat-0003021}.  Another one is the development
of chirally-improved Dirac operators such as the Neuberger operator
\cite{heplat-9806025,jansen-wup-1999} which requires the calculation
of the inverse square root of a matrix.  Algorithms for this problem
are still under development, but as they also employ iterative
solvers, it can be hoped that multigrid operators can aid in improving
performance there.  Multigrid methods have also been cited as a
possible improvement to the domain-wall formulation of lattice gauge
theory on a five-dimensional lattice \cite{heplat-0007003}.

The layout of the paper is as follows: In section 2, we introduce the
lattice Klein-Gordon and Wilson-Dirac operator and their
interpretation as diffusion on a disordered lattice, and in section 3,
the general algebraic multigrid procedure.  The numerical procedure to
construct multigrid operators is discussed in section 4, and some
applications to preconditioning and relaxation algorithms are
discussed in section 5.

\section{Model}

The model considered here is the Euclidean lattice Klein-Gordon and
Dirac operators with a $U(1)$ gauge field, discretized on a square
lattice of lattice spacing $a=1$. The formalism will also apply to
other gauge groups and can be generalized to other operators and other
types of disorder. Numerical calculations are performed in two
dimensions and on relatively small lattices ($16\times 16$), so that
the complete spectrum of the operators can be analyzed.


\subsection{Klein-Gordon operator \label{sec:KleinGordon}}

For simplicity, we start with the Klein-Gordon operator
which, without a gauge field, is conventionally written as
\begin{equation}
  \label{eq:29}
  - \Delta(x,y) + m^2 \delta_{x,y}
\end{equation}
with the next-neighbor
lattice Laplacian $\Delta(x,y)$ and the (bare) mass $m$. Its
eigenmodes are discrete Fourier modes, and its Green's functions decay 
exponentially as $e^{-m |x-y|}$ 
with the inverse correlation length $m$. Gauge disorder
is introduced as a $U(1)$ gauge phase $U_\mu(x)$ associated with the
links 
$(x,\mu)$ of the 
lattice. The disordered Klein-Gordon operator can be rewritten, up to a
normalizing factor, as
\begin{eqnarray}
  M(x,y) &=& \delta_{x,y}- \kappa Q(x,y) \quad, \nnm\\
  Q(x,y) &=& \sum_{\mu=1}^{\pm D}
                \delta_{y,x+e_\mu} U_\mu(x)  \quad.      \label{eq:1}
\end{eqnarray}
Here $x$, $y$ are sites on the lattice, $\mu$ labels positive and
negative directions $\pm 1,\ldots,\pm D$, $e_\mu$ is the unit vector
in direction $\mu$, and we use the convention $e_{-\mu} = -e_\mu$ and
$U_{-\mu}(x) = U^\conj_\mu(x - e_\mu)$.  The completely hopping
parameter $\kappa$ are borrowed from fermion terminology and
related to the bare mass $m$ in (\ref{eq:29}) by
\begin{equation}
  \label{eq:34}
  \kappa = \frac{1}{m^2 + 2D} \quad.
\end{equation}
It determines the spectrum of $M$ and thus the physical properties of
the theory. For the free theory, i.e.~$U_\mu(x) \equiv 1$, the
eigenvalues of $Q$ lie between $-2D$ (checkerboard configuration) and
$2D$ (completely constant configuration). Consequently, the
eigenvalues $\lambda_n$ of $M$ lie in a band of width $2D\kappa$
around $1$:
\begin{equation}
  \label{eq:91}
  1 - 2D \kappa \le \lambda_n \le 1 + 2D\kappa \quad.
\end{equation}
For values of $\kappa$ below a critical value $1/2D$,
\begin{equation}
  \label{eq:36}
  \kappa < \frac{1}{2D} = \kappa_{\rm crit}
\end{equation}
the spectrum of $M$ is strictly positive and $M$ invertible. The
situation for a theory in a gauge field will be discussed in the next
section. 

Physically, the interesting quantity in $M$ is the location of the
lowest eigenvalue that determines the mass gap of the theory and the
exponential decay of its Green's functions, i.e.~the correlation
functions or propagators. As $\kappa$ approaches $\kappa_{\rm crit}$,
the mass gap becomes smaller and the correlation length increases,
until at $\kappa = \kappa_{\rm crit}$, a zero eigenvalue appears and
the correlation functions, after proper subtraction, decays
logarithmically or rationally, depending on the dimension $D$.

The performance of iterative solvers for the inverse of
$M$ is linked to the condition number, i.e.~the ratio of the largest
to the smallest eigenvalue, that diverges as $\kappa \to
\kappa_c$. Unfortunately, this is exactly the region that is of
interest in lattice gauge theories, where quarks are much lighter
than the typical scales of the theory.

\subsection{Diffusion representation of $M^{-1}$\label{sec:diffusion}}

Conventionally, eq.~(\ref{eq:1}) is interpreted in lattice field theory
in terms of its spectrum that determines the dispersion relation and
thus the mass of the physical particle.  However, this description is
less useful in the presence of a gauge field, as there is no
simple expression for the eigenmodes and eigenvalues of $M$.
Alternatively, $M$ can also be interpreted as describing the diffusion
of noninteracting random walkers, i.e.~Brownian motion,
on the lattice. This can most easily
be seen by using the von Neumann series to calculate $M^{-1}$:
\begin{eqnarray}
\label{eq:31}
(1-\kappa Q)^{-1} &=& 1 + \kappa Q + (\kappa Q)^2 + (\kappa Q)^3 + 
  \ldots 
\nnm \\ (1-\kappa Q)_{xy}^{-1} &=&
\sum_{n=0}^\infty 
\sum_{\mu_1}^{\pm D} \ldots \sum_{\mu_n}^{\pm D} \, 
  \delta_{y,x+e_{\mu_1}+ \cdots \, 
  +e_{\mu_n}} \, \kappa^n \, {\cal P}(U;x,\mu_1,\ldots,\mu_n) \quad.
\end{eqnarray}
The sum here runs over all connected paths leading from $x$ to $y$, and
each path carries a gauge phase assembled from the links it traverses 
of the lattice:
\begin{eqnarray}
  \label{eq:38}
  {\cal P}_{\mu_1,\ldots,\mu_m}(U;x)
  &= & {\cal P}(U;x,x+e_{\mu_1},\dots,x+e_{\mu_1}+\ldots+e_{\mu_{n}})
  \\
  \label{eq:38b}
   &=& 
   U_{\mu_1}(x) U_{\mu_2}(x + e_{\mu_1}) \ldots
   U_{\mu_n}(x+e_{\mu_1}+\cdots+e_{\mu_{n-1}}) \quad.
\end{eqnarray}
The diffusion interpretation is based of the fact that the sum in
eq.~(\ref{eq:31}) can be generated by a random walk starting at $x$
and making random moves along the links of the lattice until reaching
site $y$.
Since the probability for each path of length $n$ is $(2D)^{-n}$,
eq.~(\ref{eq:31}) can be rewritten as an ensemble average over
random walkers:
\begin{equation}
  \label{eq:94}
  (1-\kappa Q)^{-1}_{xy} = \sum_n \left<
  \left(\frac{\kappa}{2D}\right)^n \, {\cal P}(U;x,\mu_1,\ldots,\mu_n) 
  \, \delta_{y,x+e_{\mu_1}+ \cdots +e_{\mu_n}}
  \right>_n
\end{equation}
where the average is over all random paths of length $n$, and each
random walker carries a phase factor.

In the free case, and for $\kappa=2D$, eq.~(\ref{eq:94}) is just the
time-integrated probability, or average time spent, at site $y$ for a
random walker starting out at $x$. Other values of $\kappa$ correspond
to spontaneous creation or absorption of walkers.  In the presence of
gauge disorder, each walker additionally carries a phase factor ${\cal
  P}(U)$ picked up from the links it has travesed.  Their
contributions can therefore interfere destructively, leading to a
stronger decay of the Green's function than in the free case:
\begin{equation}
  \label{eq:95}
  \left|(1-\kappa Q(U))^{-1}_{xy}\right| \le 
  \left|(1-\kappa Q(U=0))^{-1}_{xy}\right| \quad.
\end{equation}
In particular, this means that for those $\kappa$ for which the free
correlation function exists, i.e.~for $\kappa < 1/2D$, the disordered
correlation function also converges, and the critical hopping
parameter $\kappa_{\rm crit}$ of the disordered theory therefore must
be equal or larger than $1/2D$. From the discussion in the preceding
section, this reflects an upward shift of the lowest eigenvalue of $M$
that is interpreted as a dynamically generated mass. 

The value of the dynamical mass depends on the correlation
characteristics of the gauge field. If the gauge field is strongly
correlated, similar paths will have similar phase factors, the
destructive interference will be less severe and the dynamical mass
smaller than for more disordered fields. In this way, the diffusion
representation provides a qualitative picture of the spectrum of $M$
in a disordered field.

\subsection{Wilson-Dirac operator}

The Dirac operator describes the fermionic quark fields in lattice
gauge theory simulations. These fields possess an internal spin degree
of freedom and are described by a $\lfloor d/2 \rfloor$-component
spinor field. The Dirac equation is a first-order differential
equation, and its naive discretization on the lattice suffers from
fermion doubling, which is removed in the Wilson-Dirac discretization:
\begin{eqnarray}
  \label{eq:59}
  M_D(x,y) &=& \delta_{x,y}- \kappa Q_D(x,y) \quad,\nnm\\
  Q_D(x,y) &=& \sum_{\mu=1}^{\pm D}
                \delta_{y,x+e_\mu} 
                \left( 1 - \gamma_\mu \right)
                 U_\mu(x)             \quad.
\end{eqnarray}
where in two dimensions the Dirac matrices $\gamma_\mu$ are given by
\begin{equation}
  \label{eq:60}
  \gamma_1 = 
  \left( \begin{array}{cc}  1 & 0 \\ 0 & -1  \end{array} \right)
  \quad,\quad
  \gamma_2 = 
  \left( \begin{array}{cc}  0 & 1 \\ 1 & 0  \end{array} \right) \quad.
\end{equation}
Together with the unit matrix and the matrix $\gamma_5 = i \gamma_1
\gamma_2$, they form a basis for the linear operators in spinor
space. The Wilson-Dirac operator is a non-Hermitian operator and thus
has in general complex eigenvalues and different left and right
eigenvectors, but a Hermitian Wilson-Dirac operator can by defined by
multiplying it with $\gamma_5$:
\begin{equation}
  \label{eq:61}
  M_{HD} = \gamma_5 M_D \quad.
\end{equation}
For the matter of inversion, this operator is completely equivalent to
the original Wilson-Dirac operator. Due to the two-component spinors
it acts upon, it has twice as many degrees of freedom as the
Klein-Gordon operator. These additional degrees of freedom show up in
the spectrum of $M_{HD}$ as a band of negative eigenvalues mirroring
the positive band seen in the Klein-Gordon operator. The argument
used above about the relation of $\kappa$ to the low-lying
eigenvectors and the mass gap remain valid for the Dirac operator.

\section{Algebraic multigrid}


\subsection{Schur complement}

We consider the solution of a linear system
\begin{equation}
  \sum_y M(x,y) f(x) = a(y)
  \label{eq:2}
\end{equation}
with the right-hand side $a$ given and $f$ unknown. 
In the algebraic multigrid approach, the sites of the lattice ${\cal L}$ are
decomposed into a coarse lattice ${\cal L}_1$ and a fine complement lattice 
${\cal L}_2 = {\cal L} \backslash {\cal L}_1$. This decomposition can
in principle be performed without reference to the geometric locations 
of the sites. In particular, the coarse-lattice sites are not block
averages, but retain the same values as they had on the fine lattice.
Eq.~(\ref{eq:2}) can then be rewritten as
\begin{equation}
  \label{eq:3}
  \left(\begin{array}{cc} M_{11} & M_{12} \\ M_{21} & M_{22} \end{array}
  \right) \,
  \left( \begin{array}{c} f_1 \\ f_2 \end{array} \right)
  =
  \left( \begin{array}{c} a_1 \\ a_2 \end{array} \right)
\end{equation}
where the column vectors are reordered so that $f_1$ and $a_1$ are
defined on the coarse lattice, and $f_2$, $a_2$ on the fine
lattice. Applying Gaussian elimination, the fine-lattice field
$f_2$ is eliminated completely from the first row:
\begin{eqnarray}
  \label{eq:4}
  (M_{11} - M_{12} M_{22}^{-1} M_{21}) \, f_1 &=&
  (a_1 - M_{12} M_{22}^{-1} a_2) \\
  \label{eq:5}
  M_{22} \, f_2 &=& ( a_2 - M_{21} f_1 )
\end{eqnarray}
The matrix operator in front of $f_1$ plays the role of an effective
operator on the coarse lattice. It is called the {\em Schur
complement} $S_M$ of the matrix $M$ with respect to the decomposition
${\cal L}_1$:
\begin{equation}
  \label{eq:6}
  S_M = M_{11} - M_{12} M_{22}^{-1} M_{21} \quad.
\end{equation}
The simplest example of a
Schur complement is the odd-even decomposition of a square matrix with
next-neighbor interactions, where $M_{22}$ is diagonal.

\begin{figure}[htbp]
\begin{center}
\includegraphics[width=.5\hsize]{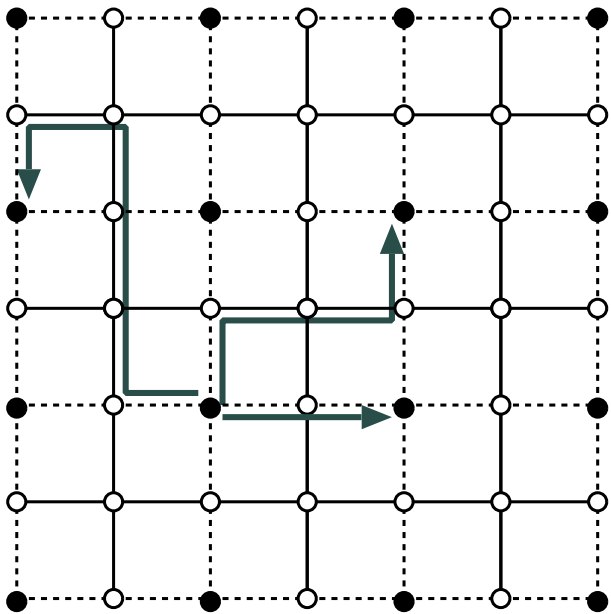}
\end{center}
    \caption{\label{fig:9}
      Two-grid decomposition of a two-dimensional lattice. Filled
      points are lattice sites of the coarse grid (${\cal L}_1$), open 
      points sites of the fine grid (${\cal L}_2$). Dashed lines
      are links that enter in $M_{12}$ and $M_{21}$, while solid lines 
      are those in $M_{22}$. The arrows show three possible
      contributions of order $2$,
      $4$, and $6$ to a general coarse-grid operator of the form (\ref{eq:19}).
    }
\end{figure}

Since the Schur complement contains the inverse of the fine-grid
submatrix $M_{22}$, algebraic multigrid must strive to find
decompositions that make $M_{22}$ as dominated by the diagonal as
possible so that a good approximation to its inverse can be found.
The method thus provides an immediate heuristic for choosing a
multigrid decomposition based only on the content of the matrix $M$
which is advantageous e.g.~for irregular discretizations.  In the case
of lattice gauge theories, the coefficients fluctuate stochastically
on the gauge group, but have the same magnitude. We thus choose a
regular decomposition where ${\cal L}_1 = (2\setZ)^D$, i.e.~the coarse
lattice is the lattice of points with all-even coordinates (cf.\ Fig.\ 
\ref{fig:9}; for a different choice see \cite{medeke-wup-1999}). Since
${\cal L}_1$ is not connected with respect to $M$, $M_{11}$ is
diagonal, $M_{12}$ links points of ${\cal L}_1$ to the remaining
points, and $M_{22}$ forms a connected lattice on the fine grid.

\subsection{Block $LU$ factorization}

Eq.~(\ref{eq:4}) and (\ref{eq:5}) can be expressed in matrix form as
an {\em block $LU$ decomposition} of $M$. Its general form is
\begin{equation}
  \label{eq:40}
  M = 
  \left(
    \begin{array}{cc}
      1 & -Q^\conj \\ 0 & 1 
    \end{array}
    \right)
    \,
  \left(
    \begin{array}{cc}
      S & 0 \\ 0 & T
    \end{array}
    \right)
    \,
  \left(
    \begin{array}{cc}
      1 & 0 \\ -Q & 1
    \end{array}
    \right) \quad.
\end{equation}
The advantage of the representation (\ref{eq:40}) is
that its inverse can be readily written as
\begin{equation}
  \label{eq:44}
  M^{-1} = 
  \left(
    \begin{array}{cc}
      1 & 0 \\ Q & 1
    \end{array}
    \right) \,
  \left(
    \begin{array}{cc}
      S^{-1} & 0 \\ 0 & T^{-1}
    \end{array}
    \right)
    \,
  \left(
    \begin{array}{cc}
      1 & Q^\conj \\ 0 & 1 
    \end{array}
    \right)
    \,
\end{equation}
requiring only the inversion of the smaller matrices $S$ and $Q$. In
multigrid terminology, applying the factorization (\ref{eq:44}) to a
vector amounts to first applying an {\em restriction operator}
$Q^\conj$ that moves information from the fine grid to the
coarse grid, then solving the {\em effective coarse-grid operator} $S$
along with the residual fine-grid operator $T$, and finally
using the {\em interpolation operator} $Q$ to move information back
from the coarse grid to the full grid.

Inserting (\ref{eq:40}) into (\ref{eq:3}) leads to the following
relations that define $S$, $Q$, and $T$.
\begin{eqnarray}
  \label{eq:41}
  S f_1 &=& a_1 - Q^\conj \, a_2 \quad \\
  \label{eq:41b}
  T \left( -Q f_1 + f_2 \right) &=& a_2 \quad.
\end{eqnarray}
Comparing with eq.~(\ref{eq:4}) and (\ref{eq:5}) gives the solution
\begin{eqnarray}
  \label{eq:44b}
  S &=& S_M = M_{11} - M_{12} M_{22}^{-1} M_{21} \quad, \nnm \\
  Q &=& -M_{22}^{-1} M_{21} \quad,\nnm\\
  T &=& M_{22}  \quad.
\end{eqnarray}

The Schur complement is the exact effective coarse-grid operator
associated with a given decomposition of the lattice. Unlike the
geometrical approach, this decomposition is performed simply by
thinning the matrix and thus without a block averaging that might
introduce artifacts in the presence of a gauge field, and the
resulting coarsened operator is defined solely in terms of the
original matrix and thus the original disorder field. In the following
sections, we will show that this procedure can be interpreted in terms
of a noninteracting random walk leading to an explicit
representation of the effective matrix in terms
of a generalized path-dependent stencil.

\subsection{Renormalization group and projective multigrid}
The Schur complement has a close relation to the renormalization group
approach used in statistical mechanics, where the Klein-Gordon and
Dirac fields are described by a path integral with a Gaussian weight.
In this approach, the partition sum
\begin{eqnarray*}
  \label{eq:73}
  Z &=& \int {\rm d}f \, \exp\left[
    -\onehalf (f,Mf) + \onehalf (a,f) + {\rm c.c.} \right] \nnm \\
  &=& \const \cdot
    e^{{\scriptstyle\frac{1}{4}} (a, M^{-1} a)}
\end{eqnarray*}
(the integration is over all lattice field configurations $f$)
generates the inverse of $M$ by
taking the derivative with respect to the source field $a(x)$:
\begin{equation}
  \label{eq:74}
  \left.
  \frac{1}{Z} \, \frac{\partial^2}{\partial a^*(x) \, \partial a(y)}
  \right|_{a=0}
  Z = \onehalf M^{-1}(x,y) \quad.
\end{equation}

In renormalization-group inspired multigrid approach, such as proposed
by Mack \cite{mack-1988}, Brower et.~al.\ 
\cite{brower-edwards-rebbi-1991} or Vyas \cite{prd-vyas}, one
introduces new variables in the path integral by means of coarsening.
In our terminology, a coarsened lattice field $F_1$ defined on on
${\cal L}_1$ is given by
\begin{equation}
  \label{eq:80}
  F_1 = f_1 + C f_2
\end{equation}
where the coarsening matrix $C$ performs an averaging over neighboring
sites in ${\cal L}_2$. (A more general coarsening would also include
an averaging over neighboring sites from ${\cal L}_1$).  The coarsened
field $F_1$ is interpolated back to the fine lattice using an
interpolation kernel $A$ and subtracted from $f_2$ to form the
residual fine-grid field:
\begin{equation}
  \label{eq:81}
  \zeta_2 = f_2 - A F_1
\end{equation}
In this way, it is hoped that some of the dynamics of the fine
lattice ${\cal L}_2$ is moved to the coarse lattice.  The whole
transformation reads
\begin{equation}
  \label{eq:82}
  \left(\begin{array}{c} F_1 \\ \zeta_2 \end{array}\right) = 
  \left(\begin{array}{cc} 1 & 0 \\ -A & 1 \end{array}\right) \,
  \left(\begin{array}{cc} 1 & C \\ 0 & 1 \end{array}\right) \,
  \left(\begin{array}{c} f_1 \\ f_2 \end{array}\right) \quad.  
\end{equation}
As its determinant is unity, $F_1$ and $\zeta_2$ can be used as new
integration variables in the path integral. The quadratic form in the
exponential is then
\begin{eqnarray}
  \label{eq:83}
  \left( f,M f \right) &=& 
  \left(\begin{array}{c} F_1 \\ \zeta_2 \end{array} \right)^\conj
  \left(\begin{array}{cc} 1 & A^\conj \\ 0 & 1 \end{array}\right) \,
  \tilde M \,
  \left(\begin{array}{cc} 1 & 0 \\ A & 1 \end{array}\right) 
  \left(\begin{array}{c} F_1 \\ \zeta_2 \end{array} \right)^\conj 
  \\
  \label{eq:83b}
  \tilde M &=& 
  \left(\begin{array}{cc} 1 & 0 \\ -C^\conj & 1 \end{array}\right) \,
  M \,
  \left(\begin{array}{cc} 1 & -C \\ 0 & 1 \end{array}\right)
  \quad,
\end{eqnarray}
and the coupling to the external fields
\begin{eqnarray}
  \label{eq:84}
  (a,f) = 
  \left(\begin{array}{c} a_1 + A^\conj a_2 - A^\conj C^\conj a_1 \\ 
                         a_2 - C^\conj a_1 \end{array} \right)^\conj \,
  \left(\begin{array}{c} F_1 \\ \zeta_2 \end{array} \right)^\conj  \quad.
\end{eqnarray}

Algebraic multigrid does not use coarsening; the coarse lattice field
is obtained by thinning out the original lattice and thus $F_1=f_1$,
$C=0$ and $\tilde M = M$.  Then the block $LU$ decomposition (\ref{eq:40})
tells us that we can make the quadratic form (\ref{eq:83}) block
diagonal by choosing $A = -M_{22}^{-1} M_{21}$, in which case the partition
sum factorizes into a coarse-grid and a fine-grid integral:
\begin{eqnarray}
  \label{eq:75}
  Z &=& \int {\rm d}f_1 \, 
        e^{-\onehalf (f_1,S_M f_1) + \onehalf (a_1 + A^\conj a_2,f_1) + {\rm c.c.}} \cdot
        \int {\rm d}f_2 \,
        e^{-\onehalf (f_2,M_{22} f_2) + 
            \onehalf (a_2,f_2) + {\rm c.c.}} \quad.        
\end{eqnarray}
The coarse scale dynamics is obtained by taking derivatives of $Z$
with respect to $a_1$ at $a_2=0$ and is therefore completely contained
in the first integral, which is governed by $S_M$, the Schur
complement.  Thus the algebraic multigrid choice of the interpolation
kernel $A$ can be characterized as that choice that leads to a
complete decoupling of fine and coarse degrees of freedom, which makes
integrating out the fine-grid degrees of freedom $\zeta_2$ trivial.
Note that the Schur complement can be used to obtain effective
fermions such as used in the blocked fermion approach to full QCD
\cite{prd-lippert}.

If a nonvanishing averaging kernel $C$ is used, one can achieve
complete decoupling in the quadratic form by applying the Schur
complement to the matrix $\tilde M$, but now $a_1$ also couples to
$\zeta_2$ in the source term (\ref{eq:84}), and the integral over
$\zeta_2$ results in a nontrivial contribution to the partition sum.
This contribution is governed by $\tilde M_{22}^{-1}$, the
propagator of the residual modes $\zeta_2$ on the fine lattice, which
is given by
\begin{equation}
  \tilde M_{22} = M_{22} - M_{21} C - C^\conj M_{12}
                 +C^\conj M_{11} C   \quad.
\end{equation}
One therefore strives to choose an averaging kernel $C$ such that
$\tilde M_{22}^{-1}$ is as local as possible, indicating that most of
the (long-range) dynamics has been moved to the coarse-lattice field
$F_1$. 

The decoupling of coarse and fine modes was at the heart of Mack's
original multigrid proposal \cite{mack-1988}, which was inspired by
constructive quantum field theory.  It was already noticed by Brower,
Rebbi, and Vicari \cite{prd-43-1965} that minimizing the coupling
between coarse and fine modes amounts in spirit to algebraic
multigrid, and using a similar reasoning, the form of the Schur
complement is mentioned in eq.~(12) of
\cite{brower-edwards-rebbi-1991} by Brower, Edwards, and Rebbi. In
practice, in the projective multigrid method the coarsening $C$ was
chosen to localize $\tilde M_{22}^{-1}$ as much as possible, and the
interpolation $A$ to maximize the decoupling between coarse and fine
modes.  An explicit method for calculating $A$ for a given $C$ is
specified as idealized multigrid algorithm by Kalkreuter
\cite{heplat-9304004,heplat-9408013}, and it can be verified that for
a ``lazy'' coarsening kernel $C=0$, the resulting interpolation kernel
has the form (\ref{eq:44b}). The actual calculation for nontrivial $C$
is numerically infeasible as it requires the solution of a nonlinear
equation at each coarse lattice site, so instead the ground-state of a
block-truncated matrix was used for the kernels, leading to the method
known as ground-state projection multigrid. The effective coarse-grid
matrix was then approximated by a Galerkin choice using the rows of
$A$ as a basis.

Using the algebraic multigrid choice of $C=0$, we cannot optimize the
decay of $M_{22}^{-1}$. However, as there is no induced coupling of
$a_1$ to the fine lattice field $\zeta_2$, the Schur complement
provides an explicit representation of the complete coarse-lattice
dynamics. In the following, we will discuss the interpretation of this
quantity based on the diffusion representation.

\subsection{Diffusion on a multigrid}
\label{sec:diffmultigrid}

In sec.~\ref{sec:diffusion}, the diffusion representation of the
inverse of a matrix $M$ was introduced. The matrix elements of the
Schur complement $S_M$ define a similar effective diffusion process
involving only coarse lattice sites with transition probabilities and
gauge phases determined by the elements of $S_M$.  As $S_M^{-1}$ is
identical with the upper left submatrix of $M^{-1}$,
\begin{equation}
  \label{eq:96}
  S_M^{-1}(x,y) = M^{-1}(x,y) \qquad\mbox{for }x,y \in {\cal L}_1
  \quad,
\end{equation}
this diffusion process is equivalent to the original diffusion process
when origin and destination of the random walk are restricted to the
coarse lattice.  Using (\ref{eq:6}), the matrix elements of $S_M$ can
itself be written in a diffusion representation:
\begin{eqnarray}
  \label{eq:39}
  S_M(x,y) &=&
  \delta_{x,y} - \sum_n \sum_{\begin{array}{c}
      (x,x_1,\ldots,x_n,y) \\
      x_1,\ldots,x_n \in {\cal L}_2 \,
    \end{array}}
  \kappa^n \, {\cal P}(U;x,x_1,\ldots,x_n,y) \quad.
\end{eqnarray}
Since the sum comes from expanding $M_{22}^{-1}$, the inner parts
$(x_1,\ldots,x_n)$ of the paths run exclusively on the fine lattice
${\cal L}_2$; they are connected to the coarse lattice ${\cal L}_1$ at
points $x$ and $y$ through the matrix elements of $M_{12}$ and
$M_{21}$ (some of these paths are shown in Fig.~\ref{fig:9}).  The matrix
elements of the Schur complement can therefore be interpreted as
resulting from random
walks on the fine complement lattice ${\cal L}_2$, and when the Schur
complement is inverted, it defines a random walk on the
coarse lattice ${\cal L}_1$, in each of whose steps all possible paths
on the fine lattice ${\cal L}_2$ are effectively taken into account.
In this way, the Schur complement allows to ``integrate out'' the
fine degrees of freedom in the diffusion process much as
renormalization group transformations do.

\begin{figure}[htbp]
\begin{center}
\includegraphics[width=.5\hsize]{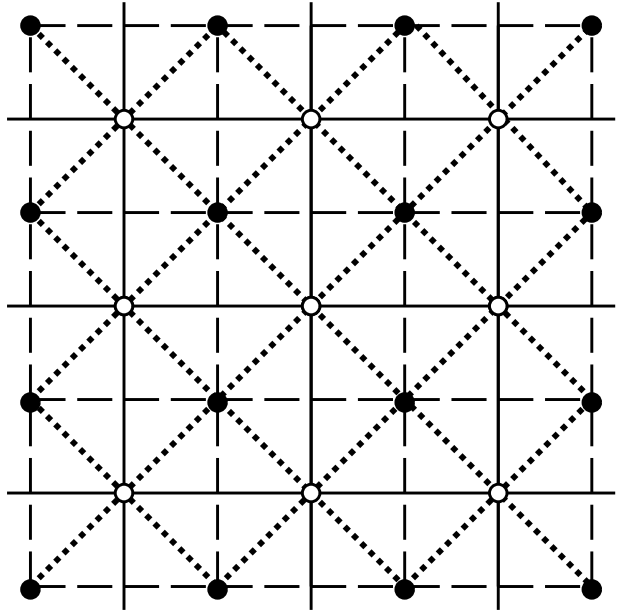}
\end{center}
    \caption{\label{fig:10}
      Two-grid decomposition of a odd-even decomposed lattice. As in
      Fig.~\ref{fig:9}, filled circles mark sites of the coarse grid,
      and open circles those of the fine grid. 
      Full lines show
      links from $M_{22}$, dashed lines those from $M_{11}$, and
      dotted lines those from $M_{12}$. $M_{11}$ and $M_{22}$ here
      form two similar grids, with $M_{12}$ connecting the two along
      the diagonals.
    }
\end{figure}

In the two-dimensional case, the situation can be made more
symmetrical by performing and even-odd-decomposition before the
multigrid decomposition. 
In an even-odd decomposition, the matrix is replaced by the effective
matrix on the even sites of the lattice,
\begin{equation}
  \label{eq:57}
  S_{ee} = 1 - M_{eo} M_{oe} \quad,
\end{equation}
where $M_{eo}$ and $M_{oe}$ are the
matrix elements of $M$ between even and odd sites, resp.  In the resulting
lattice topology, each site is linked both to its four next neighbors
in the straight direction as well to the four next neighbors in the
diagonal (cf.\ Fig.~\ref{fig:10}). Applying now the multigrid
decomposition, the lattice decomposes into two similar square
lattices. $M_{11}$ and $M_{22}$ are represented by the straight
links on each lattice, while $M_{12}$ and $M_{21}$ are carried by the
diagonals linking the two lattices. 
The effective hopping parameter on each lattice is
$\kappa_{\rm eff} = \frac{\kappa^2}{1-4 \kappa^2}$ and thus
smaller than on the original lattice, and consequently $M_{11}$ and
$M_{22}$ are less critical than $M$. In the diffusion interpretation,
even if probability was conserved on the original lattice, it is not
conserved on the two sublattices individually, as the random walkers
can move from one lattice to the other along the diagonal links
represented by $M_{12}$ and $M_{21}$.  The criticality of the system
is only regained when $M_{11}$ and $M_{22}$ are combined in the Schur
complement.

The propagator on the fine lattice $M_{22}^{-1}$ will therefore
be short-ranged and can be incorporated approximately by considering
only a limited number of hops on the fine lattice in the
expansion (\ref{eq:39}). In the following section, we exploit this to
construct an approximate Schur complement numerically.

\section{Construction of the multigrid operator}

The central problem in deriving a multigrid algorithm for disordered
systems is the construction of a suitable approximation to the Schur
complement $S_M$ and the block $LU$ decomposition
(\ref{eq:40}). 


\subsection{Numerically optimized Schur complement}

Our method is based on numerically approximating the Schur complement
as a linear combination of suitable basis operators.  
The full Schur complement $S$ usually cannot be computed exactly, 
but it can be characterized numerically by a sufficiently large 
set of pairs $\{ f^{(n)}, a^{(n)}; n=1,\ldots,N \}$ which satisfy
\begin{equation}
  \label{eq:17}
  M(U^{(n)}) f^{(n)} = a^{(n)} \quad, 
\end{equation}
where it is explicitly indicated that $M$ depends on different
realizations $U^{(n)}$ of the disorder field. 
Using the block $LU$ decomposition (\ref{eq:40}), this implies that 
$S$, $Q$, and $T$ satisfy the relations (\ref{eq:41}), (\ref{eq:41b}). 
In particular, if
the pairs are chosen such that $a_2=0$, 
their coarse-lattice components $\{ f_1^{(n)}, a_1^{(n)} \}$
characterize the Schur complement by the relation:
\begin{equation}
  \label{eq:42}
  S f_1^{(n)} = a_1^{(n)} \quad.
\end{equation}
Similarly, the interpolation operator is characterized by the pairs $\{ 
f_1^{(n)}, f_2^{(n)} \}$ using
\begin{equation}
  \label{eq:43}
  Q f_1^{(n)} = f_2^{(n)} \quad.
\end{equation}
There is no need to approximate $T = M_{22}$, as it does not contain the
inverse of $M_{22}$.
An approximation $\bar S$ to $S$, and similarly $\bar Q$ to Q, can then be
characterized with minimum bias by the mean error norm of (\ref{eq:42}):
\begin{equation}
  \label{eq:16}
  \delta^2 = \sum_n^N \left| \bar S(U^{(n)}) f_1^{(n)} - a_1^{(n)} \right|^2
\end{equation}
This quantity is an approximation to an operator norm
$|\bar S S - 1|^2$ averaged over gauge field configurations.  
The choice of pairs (\ref{eq:17}) introduces a
weight function in this norm and thus determines which functions are
well approximated by $\bar S$. Eq.~(\ref{eq:16}) can be rewritten as
\begin{equation}
  \label{eq:25}
  \delta^2 = \sum_n^N \left | \left( \bar S(U^{(n)}) S - 1 \right)
    a_1^{(n)} \right|^2
  = \sum_n^N \left | \left( \bar S(U^{(n)} - S(U^{(n)} \right)
    f_1^{(n)} \right|^2
\end{equation}
If $f_1$ is chosen randomly on the whole function space, $\delta^2$
thus approximates $|\bar S - S|^2$, while if $a_1$ is chosen randomly, it
approximates
\begin{equation}
  \label{eq:26}
  |\bar S S^{-1} -1|^2 = |(\bar S - S) S^{-1}|^2
\end{equation}
The factor $S^{-1}$ increases the weight of functions that have
eigenvalues closer to zero, and these are exactly the long-ranged
functions we are interested in. Choosing the right-hand sides $a_1$
also allows us to enforce $a_2=0$ and thus remove $Q$ from the
(\ref{eq:41}). In practice, we choose $a$ to be a delta function on a
randomly chosen coarse-lattice point, and determine $f$ by solving the
linear system (\ref{eq:17}). In other words, the $f^{(n)}$ are Green's
functions, or propagators in lattice gauge theory language, and we
look for an approximate operator $\bar S$ that well reproduces the
action of $S$ on these functions. 

If we choose a general linear combination of suitably chosen basis
operators $S^{(i)}(U)$ respecting the symmetries of the problem as
approximate operators,
\begin{equation}
  \label{eq:17a}
  \bar S(U) = \sum_i \alpha_i \bar S^{(i)}(U) \quad,
\end{equation}
the approximation error $\delta^2$ is a 
quadratic form in the coefficients $\alpha_i$:
\begin{eqnarray}
  \label{eq:18}
  \delta^2 &=&
  \sum_{ij} \alpha_i^\ast \, A_{ij} \, \alpha_j
       - \sum_i \alpha_i \, B_i^*
     - {\rm c.c.} + C \\
  \label{eq:18a}
  A_{ij} &=& \sum_n^N \, \left( S^{(i)}(U^{(n)}) f_1^{(n)}, 
                     S^{(j)}(U^{(n)}) f_1^{(n)} \right)_{{\cal L}_1}
                   \\
  \label{eq:18b}
  B_i &=& \sum_n^N \, \left (S^{(i)}(U^{(n)}) f_1^{(n)}, a_1^{(n)}\right)_{{\cal L}_1} \\
  C &=& \sum_n^N \,\left (a_1^{(n)},a_1^{(n)} \right)_{{\cal L}_1} 
\end{eqnarray}
where $(a_1,b_1)_{{\cal L}_1}$ 
denotes that scalar product on the coarse grid ${\cal
L}_1$. Thus the minimum of the approximation error and thus the 
optimal choice of coefficients is found simply by:
\begin{equation}
  \label{eq:27}
  \alpha_i = \sum_j (A^{-1})_{ij} B_j
\end{equation}
The procedure for finding the coefficients of the numerical
approximation to the Schur complement is therefore as follows:
\begin{enumerate}
\item Generate right-hand sides $a^{(n)}$ by choosing random delta functions
  on the coarse grid.
\item Find the corresponding $f^{(n)}$ by solving (\ref{eq:17}) using
  the original fine-grid operator $M$.
\item Project $a^{(n)}$ and $f^{(n)}$ to the coarse lattice.
\item Calculate the matrix elements (\ref{eq:18a}) and (\ref{eq:18b})
  and find $\alpha_i$ from (\ref{eq:27}).
\end{enumerate}
A similar procedure is used to approximate the interpolation operator
$Q$. This method is similar to Monte Carlo renormalization group as it
calculates the coefficients of the effective coarse-grid operator by
applying a block-spin transformation (in our case, simply a
projection) to an ensemble of configurations on the fine grid.  Note
that the procedure is completely general and can be used to
approximate any operator, e.g.~the Neuberger or a
renormalization-group improved operator, if its action on some sample
fields is known.


\subsection{Choice of the operator basis}
The approximate coarse-grid operator $\bar S(U)$ is a general
operator-valued function of the disorder field $U$, but the symmetries
of the problem greatly restrict the possible basis operators $\bar
S^{(i)}(U)$ from which it can be constructed.
To maintain gauge covariance, each contribution to the matrix element
$M(x,y)$ can be factored into a gauge-field independent part and a
product of the gauge field links along a connected path from $x$ to
$y$, leading to the following representation of $M$:
\begin{equation}
  \label{eq:45}
  M(x,y) = \sum_n \sum_{P = (\mu_1,\ldots,\mu_n)} \,
    m(P) \,
    \delta_{y,x + e_{\mu_1} + \cdots + e_{\mu_n}} \, 
    {\cal P}_{\mu_1,\ldots,\mu_n}(U;x)
\end{equation}
where the sum is over connected paths from $x$ to $y$, and the reduced
matrix element $m(P)$ does not depend on the gauge field $U$. 
Basis operators $S^{(i)}$ can thus be associated with the possible
relative paths $P_i = \{ \mu_1, \ldots, \mu_n \}$:
\begin{equation}
  \label{eq:19}
  \bar S^{(i)}(U)(x,y) = \Gamma^{(i)}
  \delta_{y,x + \mu_1 + \cdots + \mu_n} \, 
  {\cal P}_{\mu_1,\ldots,\mu_n}(U;x) \quad,
\end{equation}
where $\Gamma^{(i)}$ is independent of $x$, $y$, and $U$, and serves to
contain a possible internal structure of the field, e.g.~the Dirac
$\gamma$ matrices.

Other symmetries like translation, rotation, and reflection invariance
as well as hermiticity and charge conjugation generate further
restrictions on the coefficients $\alpha_i$.  In the calculations, our
computer code automatically evaluates discrete rotation and reflection
symmetry for all paths and constructs basis operators that are
invariant under these transformations. This reduces significantly the
number of degrees of freedom in (\ref{eq:18}). Hermiticity places
constraints on the real and imaginary parts of the coefficients; it is
not enforced and can be used as a check of the accuracy of the
constructed operator. The general problem of constructing generalized
Dirac operators was recently brought up in the context of
chirally-enhanced operators in \cite{heplat-0001001,heplat-0003005}
and discussed in general in \cite{heplat-0003013}.

The representation (\ref{eq:45}) has the same form as the
diffusion representation (\ref{eq:39}), with the additional
restriction that only paths moving exclusively on the fine lattice
contribute there.  The optimization process (\ref{eq:16}) thus results
in assigning weights to the different paths in the diffusion
representation that reflect how much a path contributes, on average,
to the propagator. If a complete sum over all lengths $n$ is taken,
the correct weights are of course $\kappa^{-n}$, as in the von Neumann
series. However, if the sum is truncated, the weights are modified in
the optimization process and the contributions of the individual paths
renormalized to take the obmitted paths into account.

The basis (\ref{eq:19}) can be compared to the situation without a
gauge field.  In the terminology of numerical analysis, a
translation-invariant operator is represented by a {\em stencil}
$S(a)$ using the expression corresponding to (\ref{eq:45})
\begin{equation}
  \label{eq:54}
  M(x,y) = \sum_a S(a) \, \delta_{y,x+a}
\end{equation}
where $a$ runs over the possible offsets between lattice points. The
basis operators in this case can therefore be labeled by the offset $a$,
and the equation corresponding to (\ref{eq:19}) reads
\begin{equation}
  \label{eq:55}
  S^{(i)}(x,y) = \delta_{y,x+a_i} \quad.
\end{equation}
Comparing the two cases, one can see that in the disordered case the
relative paths take over the role of the offsets, and a generalized
stencil is given by assigning each relative path a certain weight. The
optimization procedure corresponds to calculating such a generalized
stencil, much as an ordinary stencil is constructed for differential
equations.

Once the disordered stencil is known, the coarse-grid operator $S$ for
any gauge field can be constructed simply by evaluating all paths in
the set $\{S^{(i)}\}$ and combining them according to the weights
$\alpha_i$.  Note that it is not necessary to perform the construction
of the on the actual set of gauge fields, but on a representative
ensemble only, so the optimization can be performed in advance of the
simulation.

The resulting coarse-grid operator contains in particular also
contributions beyond direct neighbors on the coarse grid, similar to
improved and perfect operators that lose locality but gain a more
faithful representation of the continuum operator, and the amount of
nonlocality can be tuned by selecting the paths in the basis set. On
the other hand, if the basis set is chosen to reproduce the original
topology on the coarse lattice, the coarse-grid operator defines a new
effective disorder field on the coarse-grid. However, this disorder
field is obtained by a sum and will therefore lie outside the gauge
group, though it can be projected back to the gauge group, as is
customarily done in many renormalization schemes. In this way, the
Schur complement can be used to define a block-spin transformation for 
the gauge field.

\subsection{Construction in the diagonal subspace}
The basis (\ref{eq:19}) is the most general set of operators that can
make up a gauge-covariant operator. Its size grows exponentially with
the order of the path set and thus makes both the evaluation of the
coefficients (\ref{eq:18a}) and (\ref{eq:18b}) and the actual
construction of the approximate Schur complement very time consuming.
A smaller and thus more tractable basis can be found by considering
the operators that appear in the von Neumann series (\ref{eq:31})
which are:
\begin{eqnarray}
  \label{eq:20}
  \bar S^{(0)} &=& M_{11} \quad, \nnm\\
  \bar S^{(i+1)} &=& M_{12} M_{22}^{2i} M_{21} \quad. 
\end{eqnarray}
This basis generates exactly the same paths as the diffusion
representation (\ref{eq:19}) restricted to the fine lattice, but uses
coefficients that only depend on the length of the paths.  Obviously,
if an infinite set of basis operators is used, the optimal
coefficients are $\pm 1$ as given by the von Neumann series. However,
in a finite set, the optimization procedure amounts to finding a
polynomial approximation to the inverse function.

To see this, let $T$ be an approximation to $M_{22}^{-1}$, and write
the approximate Schur complement as
\begin{equation}
  \label{eq:22}
  \bar S = M_{11} - M_{12} T M_{21}
\end{equation}
Inserting this into (\ref{eq:16}) gives after some algebra
\begin{equation}
  \label{eq:23}
  \left| \bar S f_1 - a_1 \right|^2
  = \left( T \hat f, M_{21} M_{12} T \hat f \right)_2
    - 2 \Re \left( \hat b, T \hat f \right)_2 + \left(b, b\right)_1
\end{equation}
with the notation $(\ldots,\ldots)_2$  for the scalar product over
${\cal L}_2$ and the definitions
\begin{eqnarray}
  \label{eq:24}
  \hat f &=& M_{21} f_1 \quad,\\
   b &=& M_{11} f_1 - a_1 \quad, \nnm\\
  \hat b &=& M_{21} b \nnm\\
  \label{eq:24b}
  &=& M_{21} M_{12} M_{22}^{-1} M_{21} f_1 \quad,
\end{eqnarray}
where we used (\ref{eq:42}) and (\ref{eq:6}) in the last line.
If $T$ is a polynomial in $M_{22}$, it is diagonal in the eigenbasis
of $M_{22}$ and can be written as
\begin{eqnarray}
  \label{eq:22b}
  M_{22} &=& \sum_k \lambda_k \, v_k \otimes v_k^\conj \\
  T &=& \sum_k T_k \, v_k \otimes v_k^\conj
\end{eqnarray}
where $\lambda_k$, $v_k$ are eigenvalues and -vectors of $M_{22}$, and 
$T_k$ the diagonal elements of $T$ in the eigenrepresentation.
Inserting this into (\ref{eq:23}) gives the following
quadratic form for the
diagonal components $T_k$ of the operator $T$:
\begin{eqnarray}
  \label{eq:46}
  \left| \bar S f_1 - a_1 \right|^2 &=&
  \sum_{kl} T_k^\star \,
            ( \hat f^{\star}_k \hat f_l )
  \, \left( M_{21} M_{12} \right)_{kl} T_l
  \\ && - 2 \Re 
  \sum_k T_k \, 
         ( \hat b^{\star}_k \hat f_k )
  + \const 
\end{eqnarray}
where $\hat f_k = (v_k, \hat f)$, $\hat b_k = (v_k, \hat b)$ are the
representations of $\hat f$ and $\hat b$ in the eigenspace of
$M_{22}$, and similarly for $(M_{21} M_{12})_{kl}$. Eq.~(\ref{eq:24b})
relates these quantities by
\begin{equation}
  \label{eq:51}
  \hat b_k = \sum_l (M_{21} M_{12})_{kl} \, \lambda_l^{-1} \, 
             \hat f_l
\end{equation}
which gives the following expression for (\ref{eq:46}):
\begin{eqnarray}
  \label{eq:52}
  \left| \bar S f_1 - a_1 \right|^2 &=&
  \sum_{kl} \left( T_k - \frac{1}{\lambda_k} \right)^\star \, 
  ( \hat f^{\star}_k \hat f_l ) \,
  \left( M_{21} M_{12} \right)_{kl} \,
  \left( T_l - \frac{1}{\lambda_l} \right) + \const \nnm\\
  &=& 
  \sum_k
  \left|
    \sum_l
    \left(M_{12}\right)_{kl} \hat f_l \,
    \left( T_l - \frac{1}{\lambda_l} \right) 
  \right|^2 + \const \quad.
\end{eqnarray}
Minimizing this quantity thus amounts to finding 
a best approximation $T_k$ to
$\lambda_k^{-1}$ in a weighted square norm specified by $M_{12} \hat f
= M_{12} M_{21} f_1$. 

\begin{figure}[htbp]
\begin{center}
\includegraphics[width=.8\hsize]{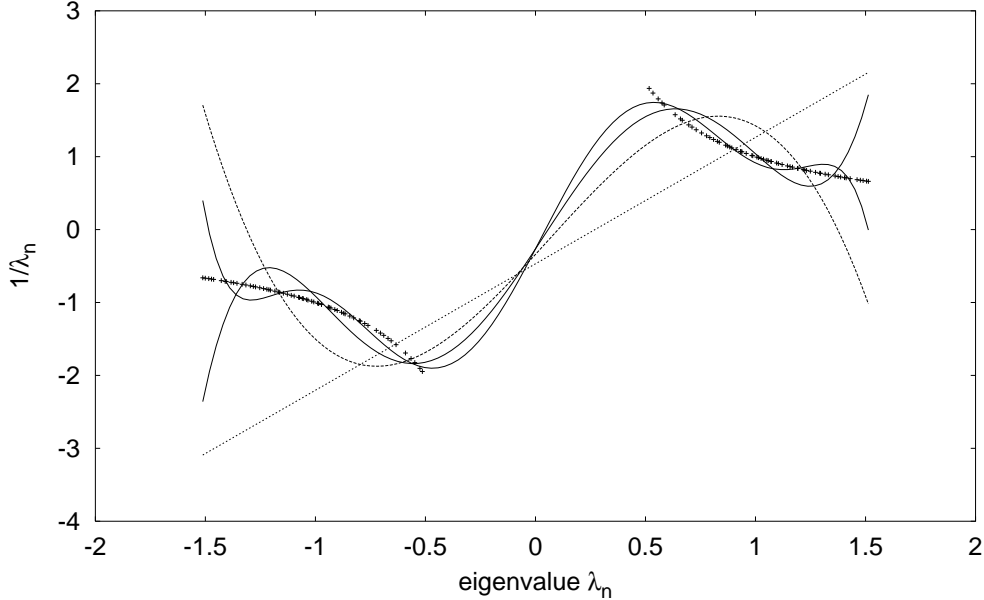}
\end{center}
    \caption{\label{fig:1}
    Numerical evaluation of the optimal polynomial 
    approximation to $M_{22}^{-1}$. The curves show the polynomial
    approximations of order 1, 3, 5, and 7 to the inverse
    function. The actual eigenvalues are marked with crosses.}
\end{figure}

Choosing for $T$ a polynomial form
\begin{equation}
  \label{eq:53}
  T = \sum_{n=0}^N \alpha_{n+1} \, M_{22}^{2n} \qquad,\quad
  T_k = \sum_{n=0}^N \alpha_{n+1} \, \lambda_k^{2n}
\end{equation}
in accordance with the expansion (\ref{eq:20}),
the approximate Schur complement then takes the form
\begin{equation}
  \label{eq:53b}
  \bar S = M_{11} 
         - \sum_{k=0}^N \alpha_{k+1} M_{12} M_{22}^{2k} M_{21}
\end{equation}
Fig.~\ref{fig:1} shows a numerical evaluation of such an approximation
using the Dirac operator on a sample $U(1)$ configuration.

An operator of the form (\ref{eq:53b}) can be computed relatively
easily: The coefficient $\bar S(x,y)$ between two sites is given by
summing the phase factors over all paths from $x$ to $y$ on ${\cal
  L}_2$
with weights
depending only on the length of the paths and given by the
coefficients $\alpha_k$.  This is a simplification over the full path
set (\ref{eq:19}), where each path can be assigned its own weight.  If
only a finite number $N$ of coefficients is used, the operator is
local, but contains more than next-neighbor interactions. Note that in
a typical iteration scheme the calculation of the matrix elements of
$\bar S$ has to be done only once, and the complexity of the iteration
depends only on the sparsity of the resulting $\bar S$.

\begin{figure}[p]
  \begin{center}
    \centerline{\hss 
      \includegraphics[width=3.8in]{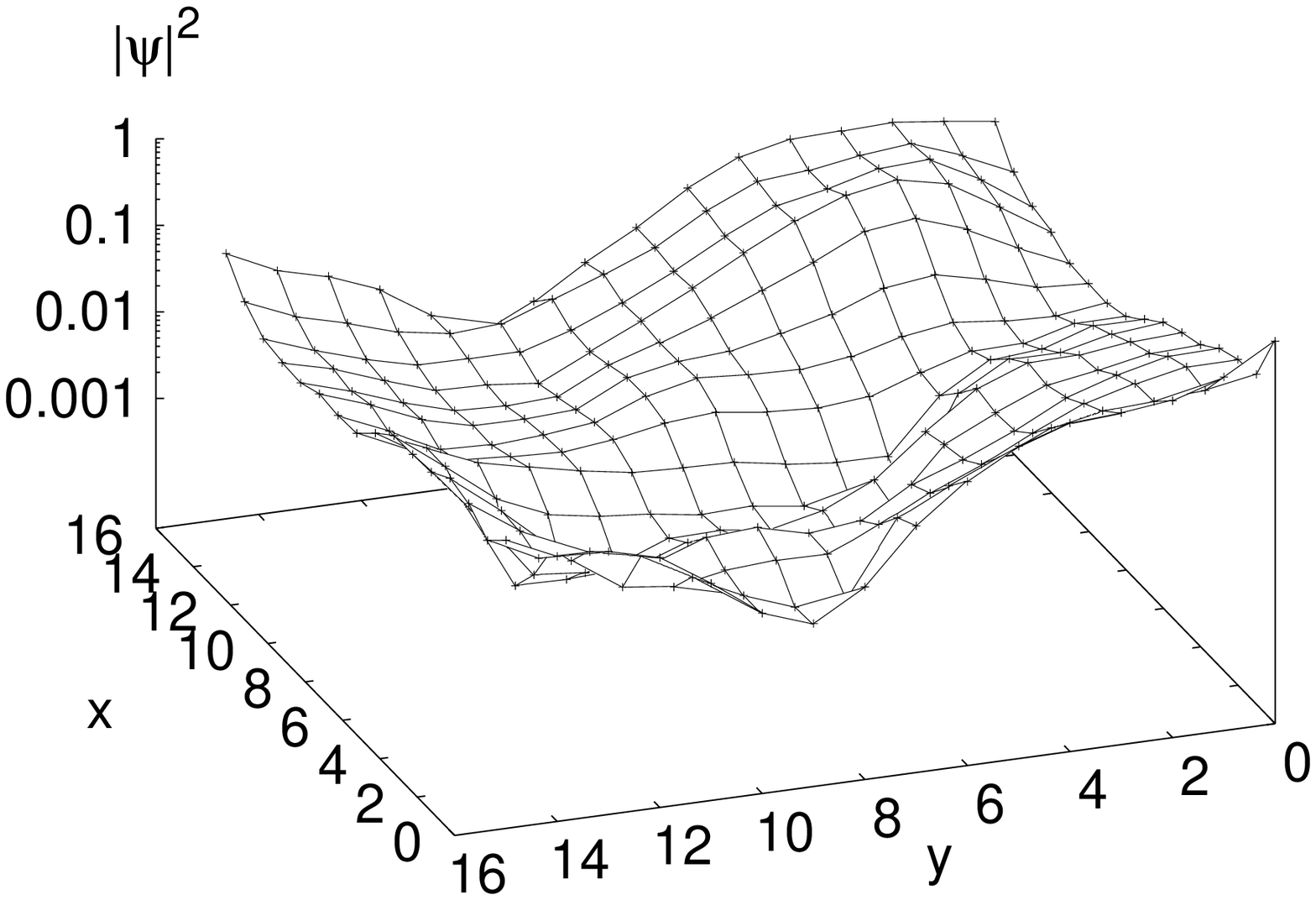} \kern-60pt
      \includegraphics[width=3.8in]{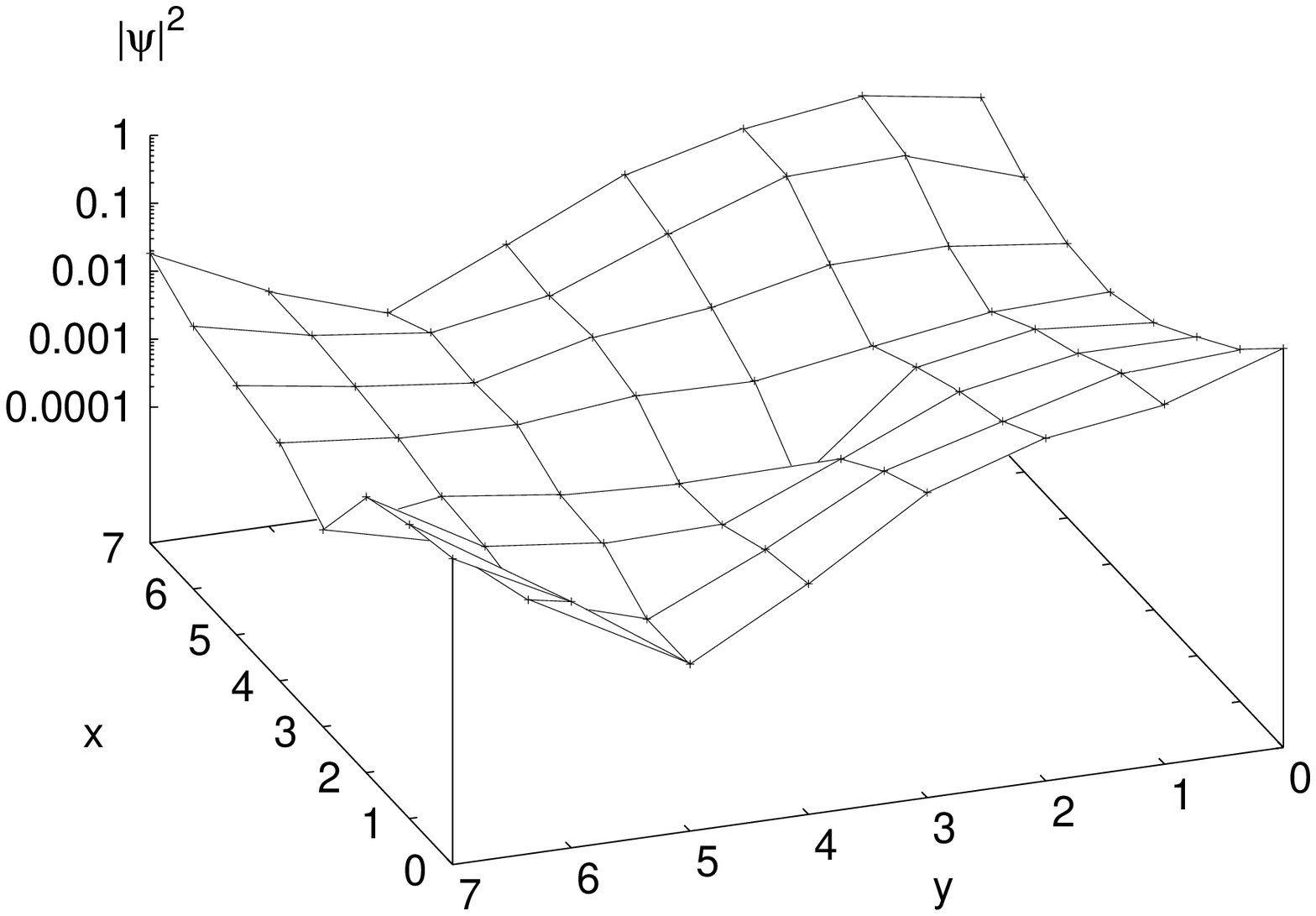} \hss}
    \caption{%
      Absolute value of a sample fermionic Green's function on an
      $16\times 16$ lattice in a $U(1)$ background field for the
      original Dirac operator (left) and the approximate coarse-grid
      operator obtained from the Schur complement. The source is
      located at the origin; since the lattice is periodic, it shows
      up at the four corners of the plot.
      }
    \label{fig:12}
  \end{center}

  \begin{center}
    \includegraphics[width=.8\hsize]{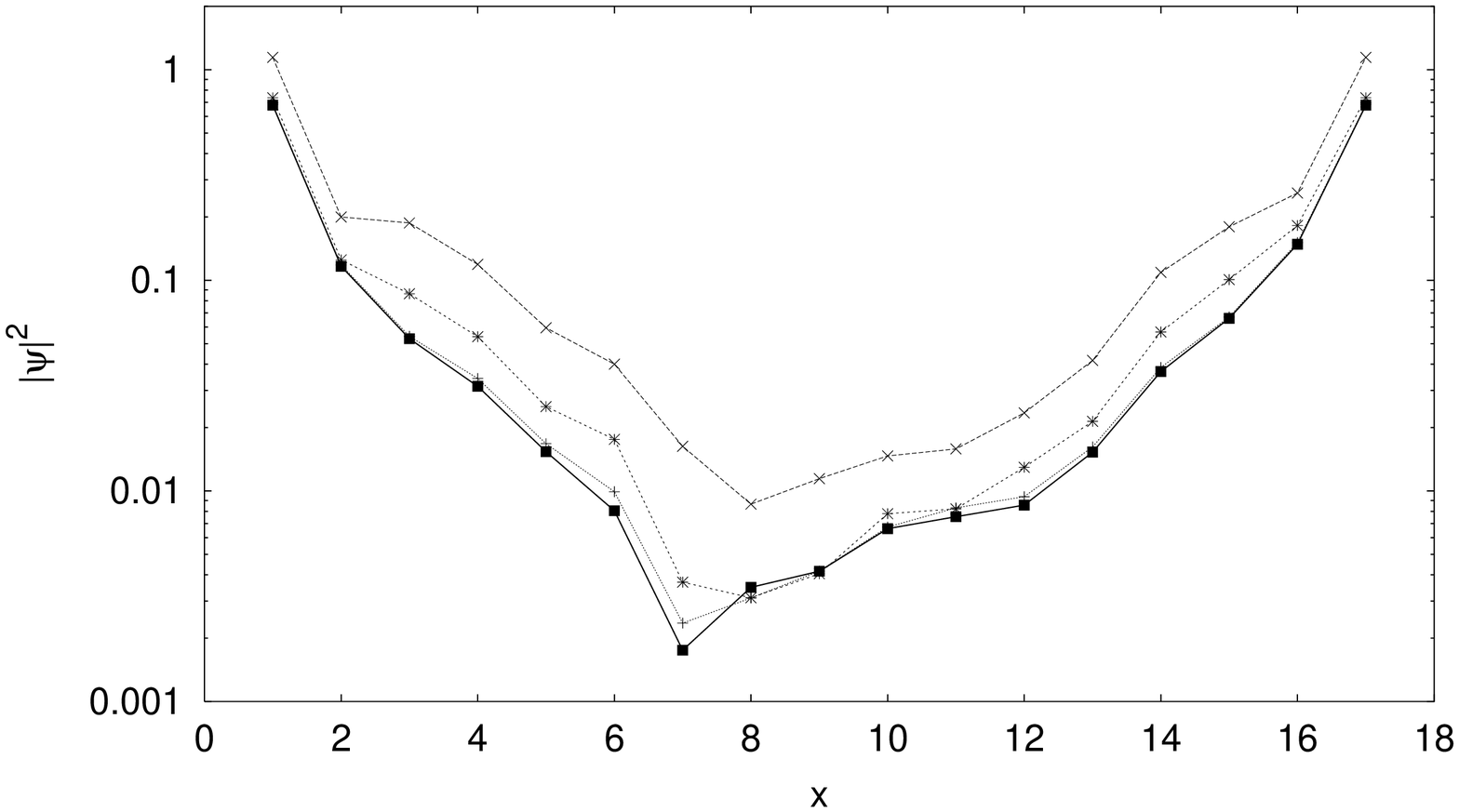}
    \caption{%
      Cut through the example fermionic Green's function (black
      squares) compared to Green's functions calculated from an
      approximate Schur complement and interpolation operator 
      of order 1, 2, and 4.
      }
    \label{fig:13}
  \end{center}
\end{figure}

\subsection{Numerical results}

We have numerically evaluated the optimal operator both for the full
path set of eq.~(\ref{eq:19}) and the subset defined in
eq.~(\ref{eq:20}). The calculations were performed for the Dirac
equation on a $16\times 16$ lattice in two dimensions in a $U(1)$
gauge field generated at $\beta=3.0$, which results in relatively
smooth configurations and long-ranged correlation
functions. Fig.~\ref{fig:12} and \ref{fig:13} show sample Green's
functions of the original Dirac matrix and the approximate Schur
complement. The disorder from the gauge field shows up clearly in the
Green's function, but the various approximations to the Schur
complement reproduce the irregularities and the overall form of the
function quite well.


\begin{figure}[p]
  \begin{center}
    \includegraphics[width=.8\hsize]{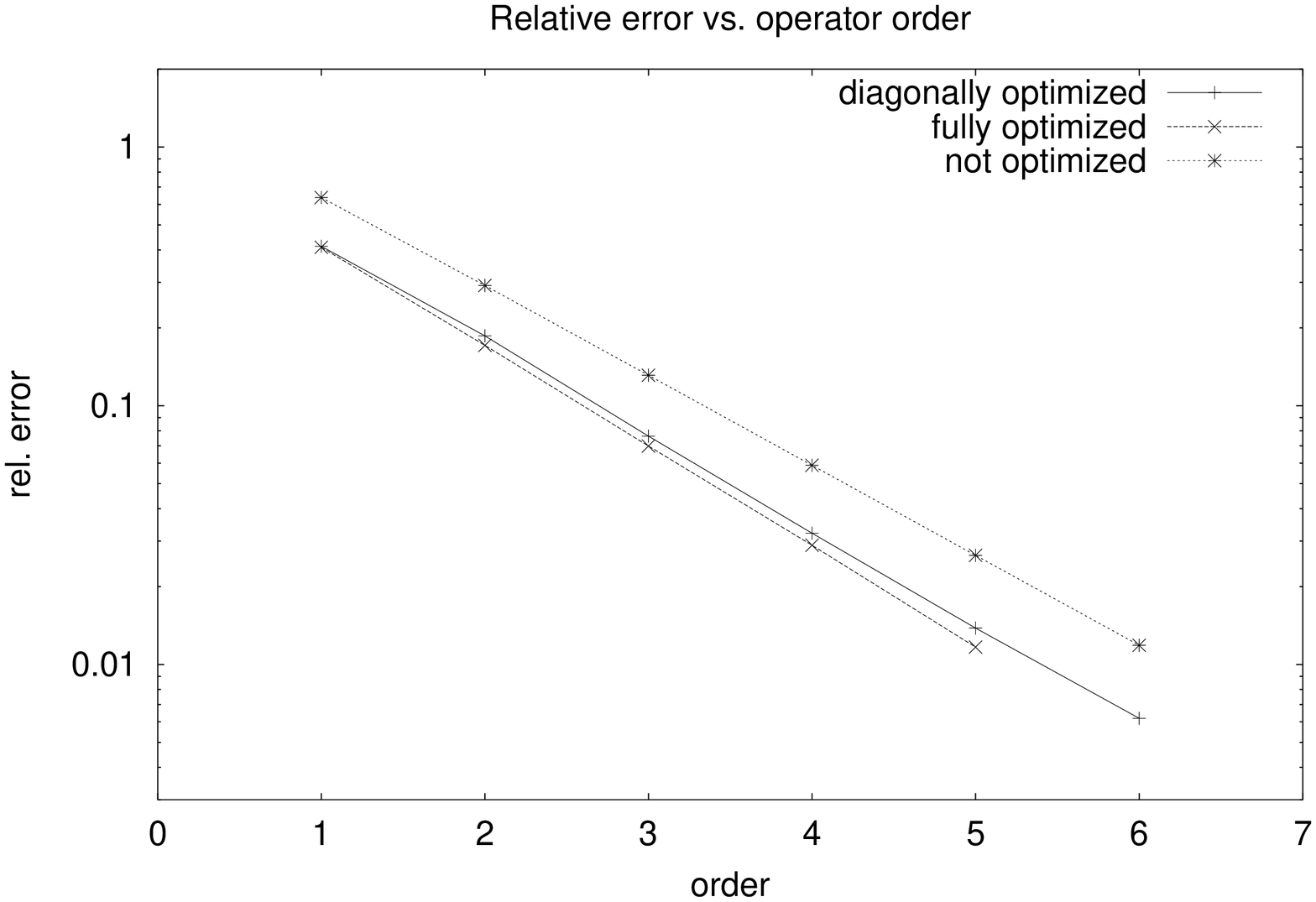}
    \caption{%
      The relative approximation error as a function of the order of the
      approximation, calculated in an ensemble of 10 independent
      $16\times16$ $U(1)$ gauge configurations at $\beta=3.0$ and
      $\kappa=0.265$. The full line gives the result from the
      optimization in the diagonal subspace, the dashed line from the
      full optimization. The dotted line is the relative error of the
      truncated von Neumann series of the same order.
      }
    \label{fig:3}
  \end{center}

  \begin{center}
    \includegraphics[width=.8\hsize]{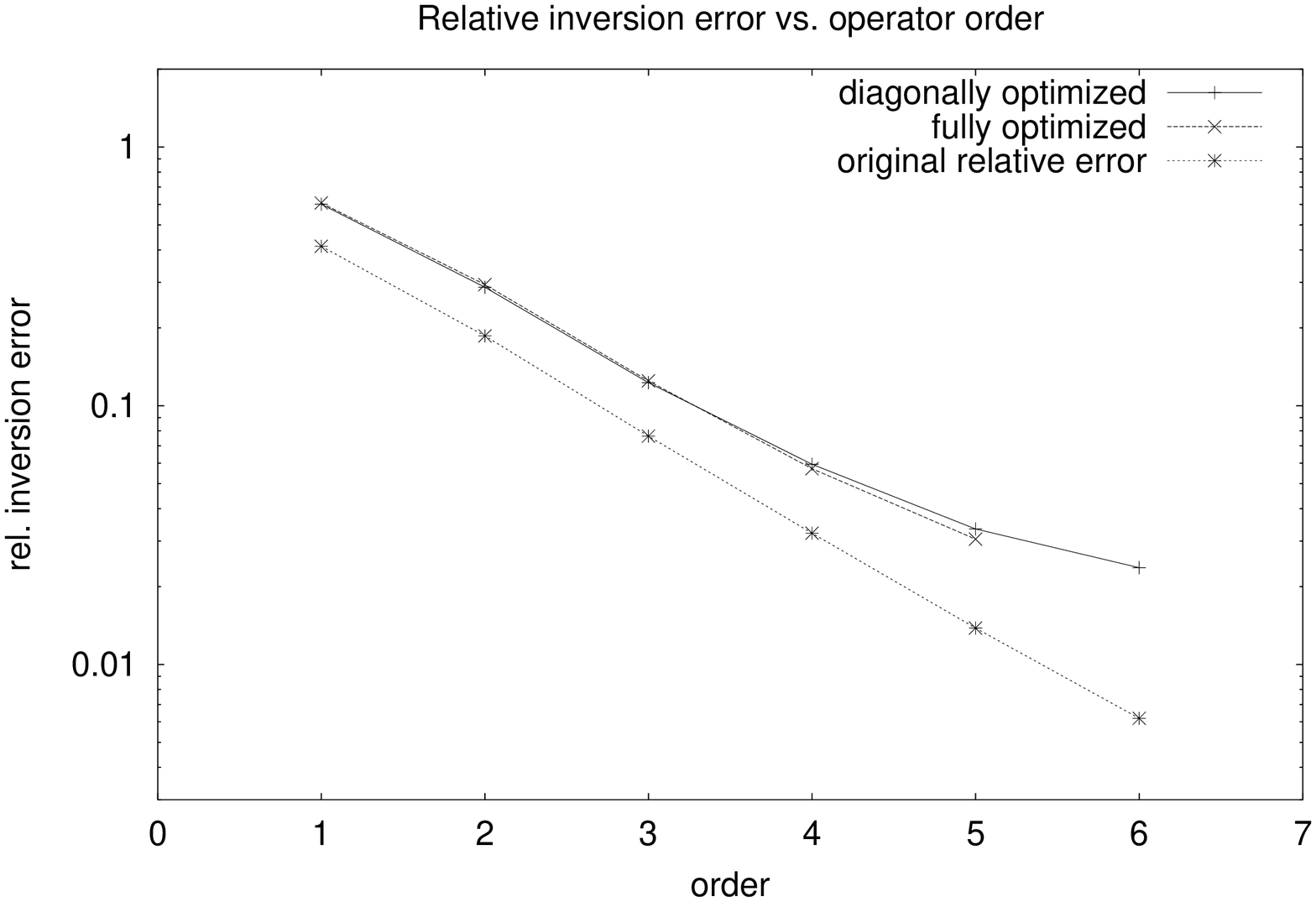}
    \caption{%
      The relative inversion error as a function of the approximation
      order for both diagonal and full optimization. For comparison, 
      the original relative error from Fig.~\ref{fig:3} is also
      shown.
      }
    \label{fig:3a}
  \end{center}
\end{figure}

This is quantified in fig.~\ref{fig:3}, which shows the average relative error
\begin{equation}
  \label{eq:70}
  \sqrt{\frac{\delta^2}{\sum_n^N |a_1^{(n)}|^2}}
\end{equation}
evaluated in an ensemble of ten sample $U(1)$ gauge configurations
generated at $\beta=3.0$ on a $16\times16$-lattice. For each
configuration, five different Green's function at $\kappa=0.265$ on each
configuration were calculated and the coefficients (\ref{eq:18a}) and
(\ref{eq:18b}) that characterize the error ellipsoid were evaluated.
For the diagonal optimization, the basis (\ref{eq:20})
with maximum order $i$ from $1$ to $6$, corresponding to a maximum
path length from $2$ to $12$, was used. For the full optimization, a complete
path basis of paths up to length $2i$ constrained to the discrete
rotation and reflection symmetry was taken. For comparison, the
relative error resulting from the von Neumann series is also shown.
The errors show a very clean logarithmic dependence on the order. As
the number of paths in the operator increases exponentially with
order, this corresponds to a polynomial dependence of the error on the
number of paths.

The full path set gives only a slightly better approximation than the
diagonal approximation, though it is significantly more difficult to
evaluate, as the number of coefficients grows exponentially. Its main
advantage is that it allows to fine-tune the number and
characteristics of the paths, e.g.~the maximum distance between end
points.  Fig.~\ref{fig:3} shows the resulting errors when the error
ellipsoid is restricted to a certain number of paths. It turns out
that the relative error has a power law dependence on the number of
paths, with an exponent of approximately $-0.55$ in this case. The
contributions from different operators are relatively uniform except
for the operators in the tails that contribute little or nothing, so
the accuracy of the approximate Schur complement can be fine-tuned by
varying the number of paths in the operator.

The relative error measures how well the approximate Schur complement
reproduces a delta function when applied to a Green's function. For
actual applications, the more relevant quantity is the relative
inversion error that looks at how good the inverse of the
approximate Schur complement reproduces the Green's function of the
original matrix. This is quantified by
\begin{equation}
  \label{eq:71}
  \delta_{\rm inv.}^2 = 
  \sum_n^N \left| \bar f_1^{(n)} - S^{-1}(U^{(n)})  a_1^{(n)}
  \right|^2
\end{equation}
with the relative error taken relative to the norm of $f_1^{(n)}$:
This quantity was evaluated numerically by computing Green's function
of the approximate Schur complement on the coarse lattice and
comparing the results to the Green's functions of the original
Wilson-Dirac matrix.  Fig.~\ref{fig:3a} shows the relative error as a
function of the approximation order both for the diagonal and the full
optimization. It also decreases exponentially but is about twice the
magnitude of the original error. 

\subsection{Spectrum}

The low eigenvalues of the original matrix $M$ have a direct relation
to those of the Schur complement $S_M$. If $f$ is an eigenvector of
$M$ with eigenvalue $\lambda$,
\begin{equation}
  \label{eq:86}
  M f = \lambda f \quad,
\end{equation}
this becomes using (\ref{eq:4})
\begin{eqnarray}
  \label{eq:87}
  S f_1 &=& \lambda f_1 - M_{21} M_{22}^{-1} \, \lambda f_2 \nnm\\
  M_{22} f_2 &=& \lambda f_2 - M_{21} f_1 \quad.
\end{eqnarray}
Eliminating $f_2$ in the first line yields a modified eigenvalue
equation for $S$:
\begin{equation}
  \label{eq:88}
  S f_1 = \lambda \, \left(
          1 + M_{12} M_{22}^{-1} (M_{22} - \lambda)^{-1} M_{21}
          \right) \, f_1 \quad.
\end{equation}
If the eigenvalues of $M_{22}$ are large, as argued in
sec.~\ref{sec:diffmultigrid}, the additional term on the right-hand
side multiplying the eigenvalue is approximately
\begin{equation}
  \label{eq:89}
  1 + M_{12} M_{22}^{-2} M_{21} = 1 + Q^\conj Q \quad.
\end{equation}
Assuming further that $M_{22}^{-1}$ is 
short-ranged, the second term can be approximated by a constant
$\alpha$ when acting on a low eigenvector, and (\ref{eq:88}) becomes
\begin{equation}
  \label{eq:90}
  S f_1 \approx \lambda (1+\alpha) f_1 \quad.
\end{equation}
This rescaling of the eigenvalues accounts for the different lattice
constant of the coarse grid.  The appearance of the interpolation
kernel $Q$ in (\ref{eq:89}) shows that it accounts for the parts
of the original eigenvector that were on the fine grid and therefore
dropped. In a geometric multigrid method, this is usually done by the
coarsening prescription which moves information from fine to coarse
degrees of freedom; in algebraic multigrid, it results directly from
the $LU$ decomposition.

\begin{figure}[htb]
  \begin{center}
    \includegraphics[width=\hsize]{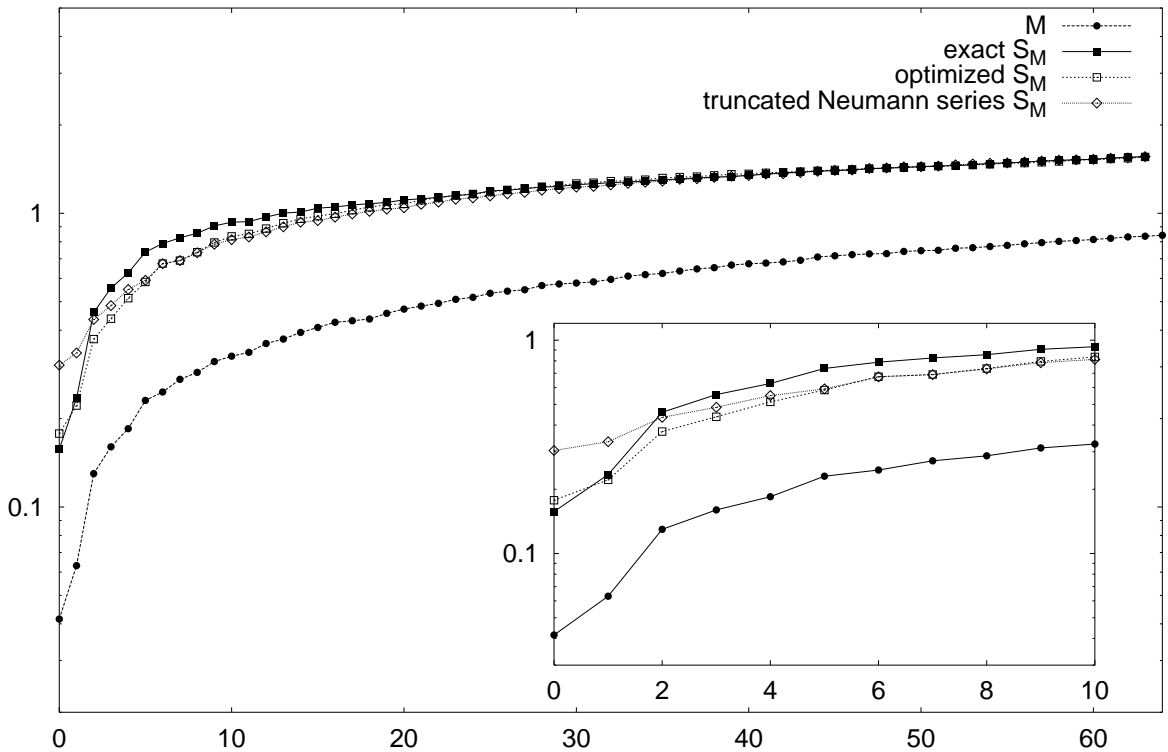}
    \caption{%
      Lowest positive eigenvalues of the effective coarse-grid
      operator $S$. The lower curve (filled circles) shows 
      eigenvalues of the full Dirac matrix on a $16\times 16$ lattice
      with a sample $U(1)$ gauge field at $\beta=3.0$. The three
      upper curves are the corresponding eigenvalues on the coarse
      grid, calculated from the exact Schur complement (full squares), 
      the order-2 optimized operator (open squares), and the order-2
      von Neumann series (open diamonds). The inset enlarges low
      eigenvalues. 
      }
    \label{fig:11}
  \end{center}
\end{figure}

Fig.~\ref{fig:11} shows the lowest positive eigenvalues of the Dirac
matrix and different forms of the Schur complement for a sample $U(1)$
gauge configuration at $\beta=3.0$. With a rescaling factor of
approximately $1+\alpha \approx 4$, the exact Schur complement $S_M$
reproduces the low eigenvalues of $M$ well, and so does the
order-2 optimized operator. The truncated von Neumann series, however,
deviates significantly from the true results for the lowest
eigenvalues, which are expected to be very important for the
convergence properties when used as a preconditioner.

\begin{figure}[p]
  \centerline{\includegraphics[width=.8\hsize]{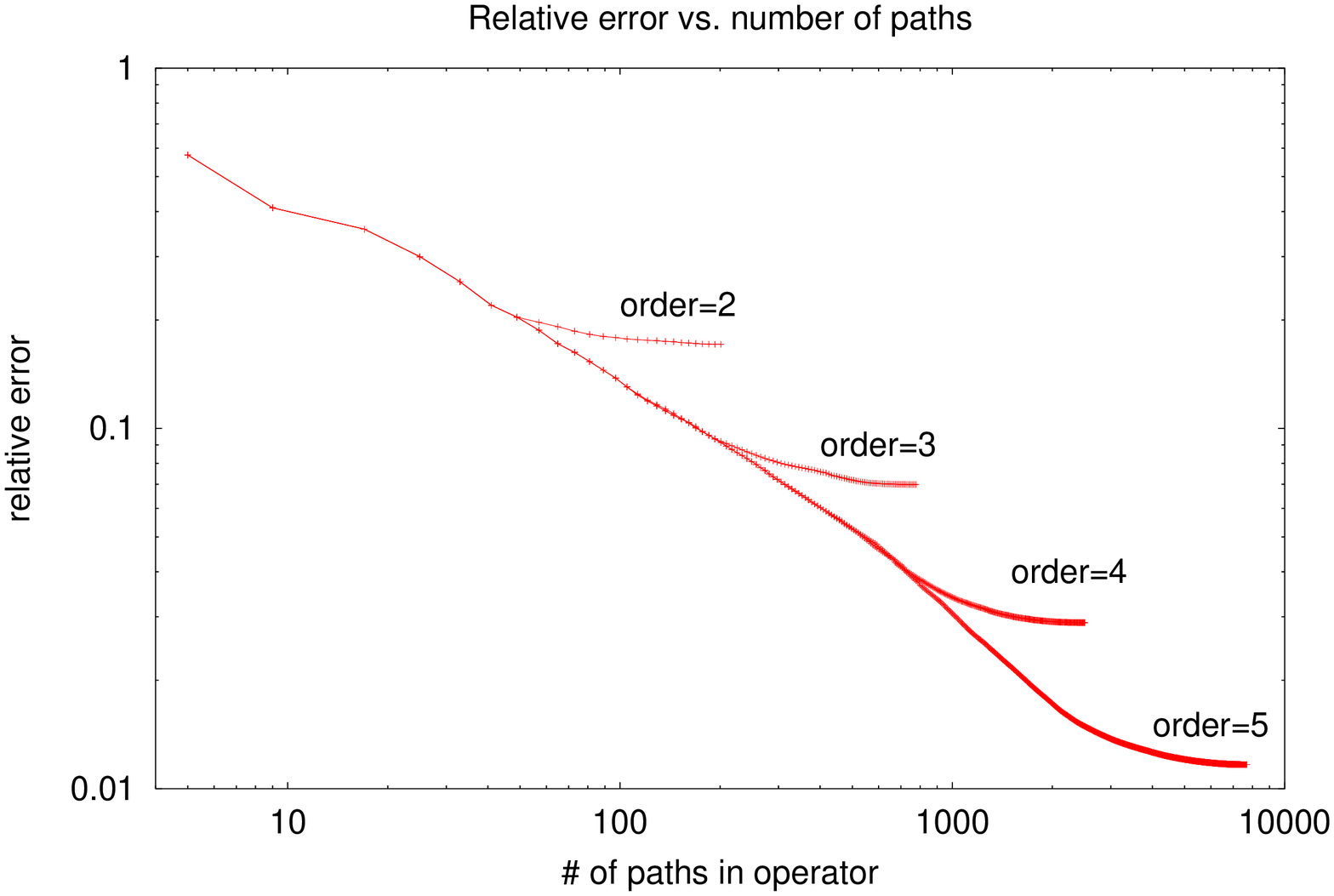}}
    \caption{The relative error as a function of the number of 
      paths in the approximate Schur complement for different orders
      of the path set. The paths have been reordered so that the
      relative error decays fastests.}
    \label{fig:2}

  \begin{center}
    \includegraphics[width=.8\hsize]{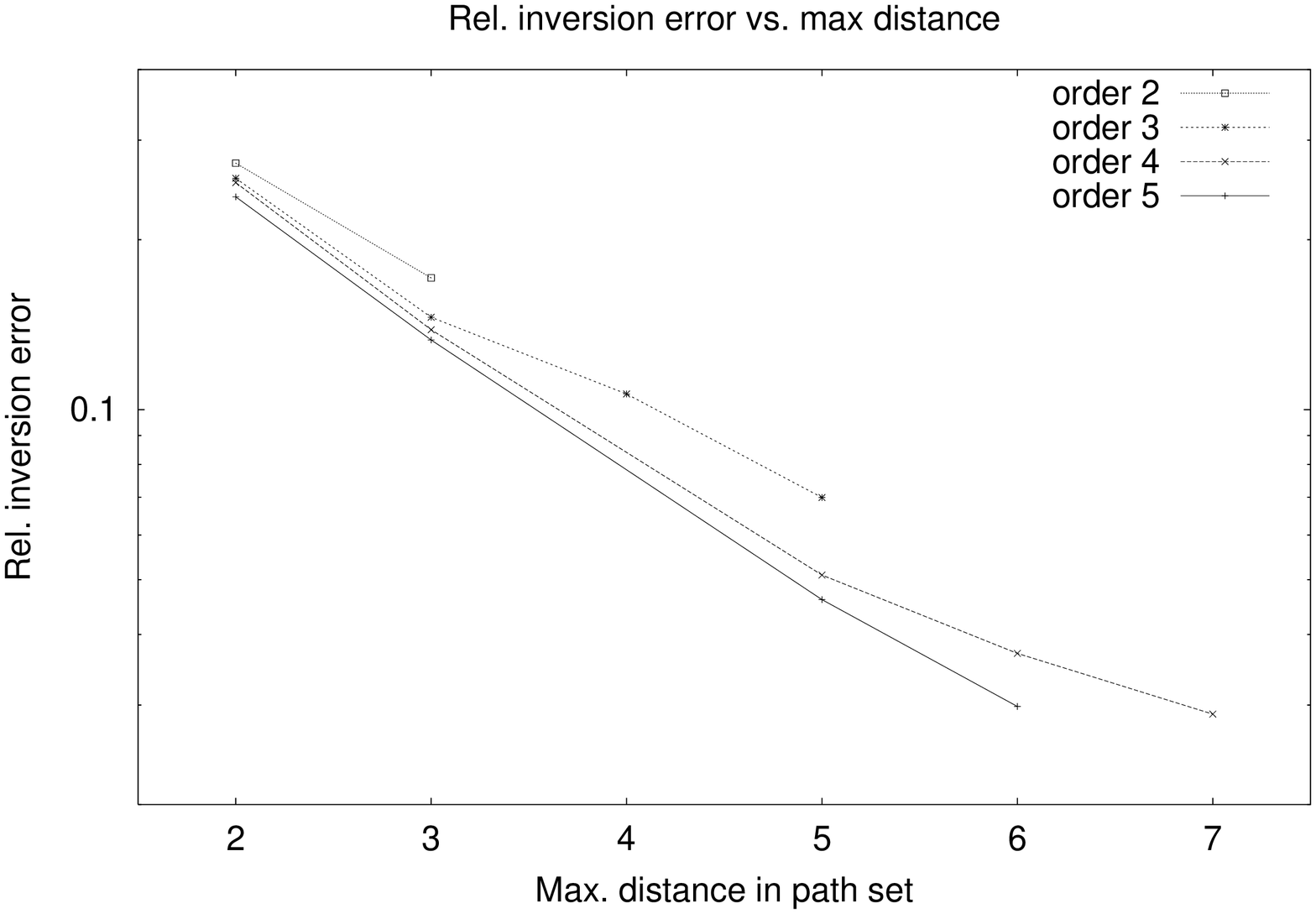}
    \caption{%
      The relative error from the full optimization as a function of
      the maximum distance of path end points in the path set, shown
      for different maximum lengths of the path.
      }
    \label{fig:4}
  \end{center}
\end{figure}

\section{Applications}

While a full discussion of the performance of multigrid operators in
different algorithms is outside of the scope of this paper, we give
here two examples of how they can be applied in numerical algorithms.
Preconditioning uses the block $LU$ decomposition provided by the
multigrid decomposition to improve the condition number and thus
convergence in iterative algorithms. Multigrid relaxation is the
classical multigrid algorithm for solving the system (\ref{eq:2}) and
allows a comparison between a two-grid iteration and the original
fine-grid iteration. In all calculations, a $16 \times 16$ lattice was 
used to allow the calculation of complete spectra. Note that only
two-grid algorithms are considered; for realistic problems, the
procedure should be repeated on several grid levels.

\begin{figure}[htbp]
  \begin{center}
    \includegraphics[width=\hsize]{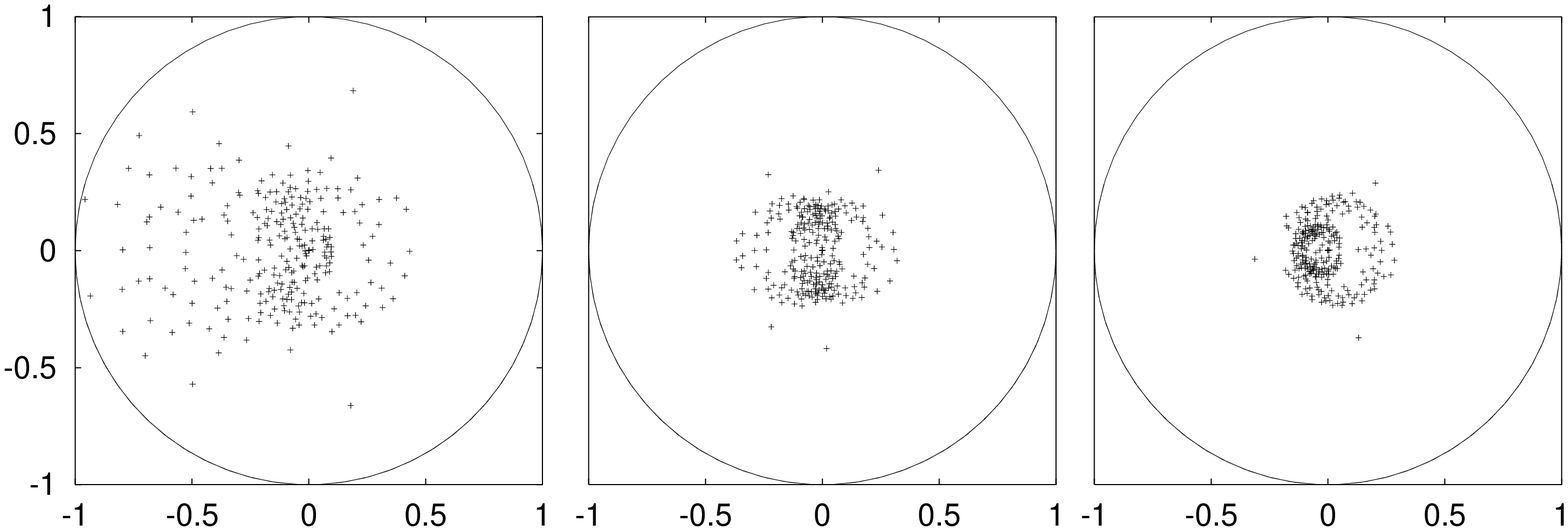}
    \includegraphics[width=\hsize]{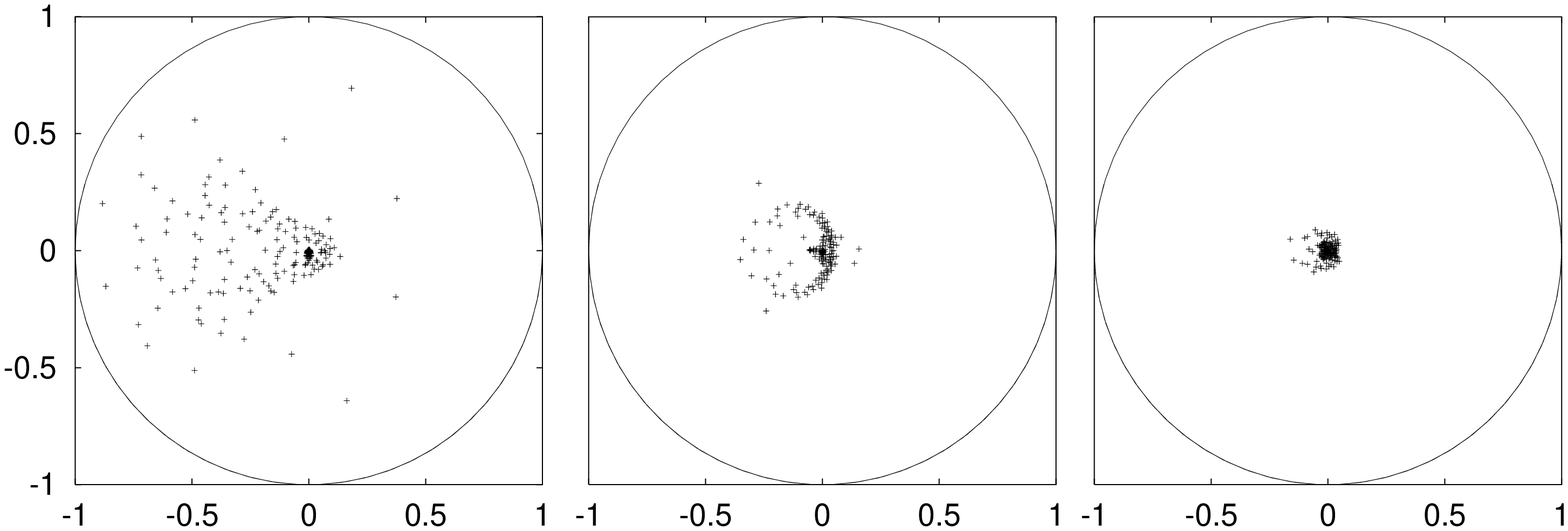}
    \caption{
      Spectra of the multigrid preconditioned Dirac matrix $1 - \bar
      M^{-1} M$ for a sample gauge field configuration.
      Optimized approximations of order 1, 2, and 3 where used from left to
      right for the Schur complement.
      The top row shows the spectra obtained using an approximate
      interpolation operator, the bottom row the spectra using the
      exact interpolation operator. Preconditioning with an exact
      Schur complement and interpolation operator would result in a
      single point at the origin.
      \label{fig:5}
    }

    \includegraphics[width=\hsize]{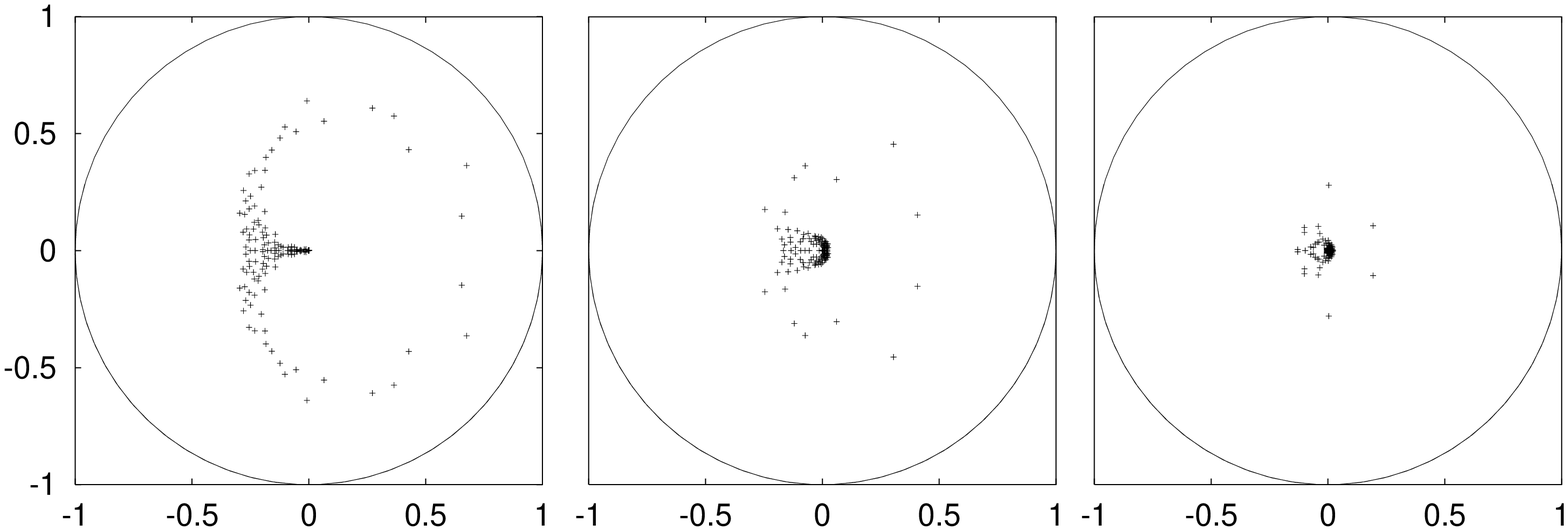}
    \caption{
      The preconditioned spectrum of fig.~\ref{fig:5},
      using the von Neumann series as approximate Schur complement.
      \label{fig:6}
    }
  \end{center}
\end{figure}

\subsection{Preconditioning}

In preconditioning, the linear system
\begin{equation}
  \label{eq:69}
  M f = a
\end{equation}
is replaced by an equivalent system
\begin{equation}
  \label{eq:69b}
  \bar M^{-1} M f = a' = \bar M^{-1} a
\end{equation}
using a preconditioning matrix $\bar M$ that is an easily invertible
approximation to $M$. If $\bar M$ is sufficiently close to $M$, the
spectrum of the preconditioned matrix $\bar M^{-1} M$ is contracted towards
unity and its condition number reduced. For multigrid preconditioning,
one uses an approximate block LU decomposition of the form
(\ref{eq:40}) as the preconditioner. The preconditioned matrix is then
\begin{eqnarray}
  \label{eq:97}
  && \bar M^{-1} M =
  \left( \begin{array}{cc} 1 & 0 \\ \bar Q & 1 \end{array}
  \right) \,
  \left( \begin{array}{cc} \bar S^{-1} & 0 \\ 0 & \bar T^{-1} 
  \end{array} \right) \,
  \left( \begin{array}{cc} 1 & \bar Q^\conj \\ 0 & 1 \end{array}
  \right) \,
  \left( \begin{array}{cc} M_{11} & M_{12} \\ M_{21} & \bar M_{22}
  \end{array} \right) 
\end{eqnarray}
If $Q$ and $T$ are chosen to be exact, $Q=-M_{22}^{-1} M_{21}$ and
$T=M_{22}$, this reduces to
\begin{equation}
  \label{eq:98}
  \bar M^{-1} M =  
  \left( \begin{array}{cc} 
    \bar S^{-1} S_M &
    0\\
    \bar Q \bar S^{-1} ( \bar S^{-1} S_M - 1 ) &
    1 \end{array} \right)
  \quad.
\end{equation}
The properties of the preconditioner are therefore determined by $1 -
\bar S^{-1} S_M$, which is the quantity that was optimized in the
optimization process (\ref{eq:26}) above.


Fig.~\ref{fig:5} shows sample spectra of $1 - \bar M^{-1} M$ with the
Schur complement approximated to order 1, 2, and 3. If an exact
Schur complement was used, the spectrum would reduce to a single point
at the origin; the radius of the disk on which the eigenvalues lie in the
complex plane is a measure of the quality of the preconditioner. The
figure shows the result for two different choices of the interpolation
matrix $Q$: In the bottom row, the exact interpolation $Q=-M_{22}^{-1}
M_{21}$ was used. Since $Q$ is, as opposed to $S$, never inverted in
the process, this is numerically legitimate and no more complex than
the application of the approximate $S^{-1}$. The top row shows the
result when an approximate interpolation matrix $\bar Q$ was used that 
was calculated in the same way as $S$ in an optimization process. It
shows that only if a high order of approximation for $S$ was chosen, it
makes sense to use the exact interpolation operator, otherwise it just 
leads to a larger concentration of eigenvalues without reducing the
radius of the disk.

For comparison, fig.~\ref{fig:6} shows the spectrum of a matrix that
was preconditioned using von Neumann series (and the exact
interpolation operator) to the same order as before. While most modes
are quite well reduced, there remain even at order 3 a few eigenvalues 
far away from the origin. These probably correspond to low eigenvalues of
the original matrix that are not properly approximated by the von
Neumann series, as seen in fig.~\ref{fig:11} above.

\begin{figure}[p!]
  \begin{center}
    \includegraphics[width=\hsize]{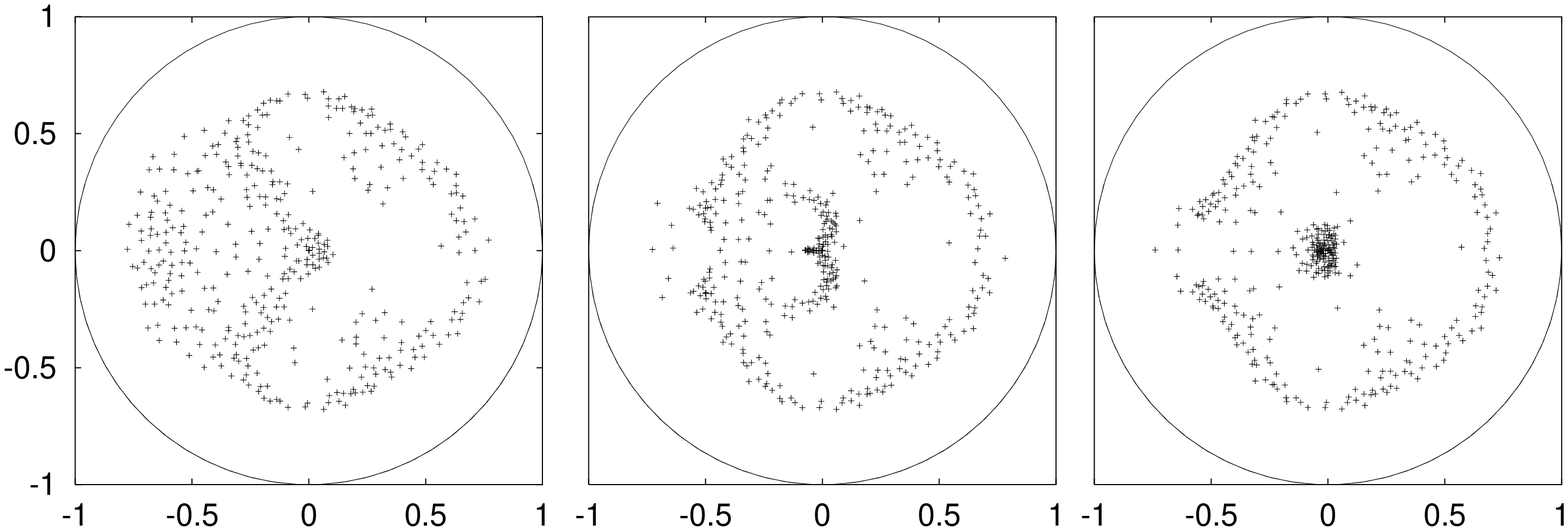}
    \includegraphics[width=\hsize]{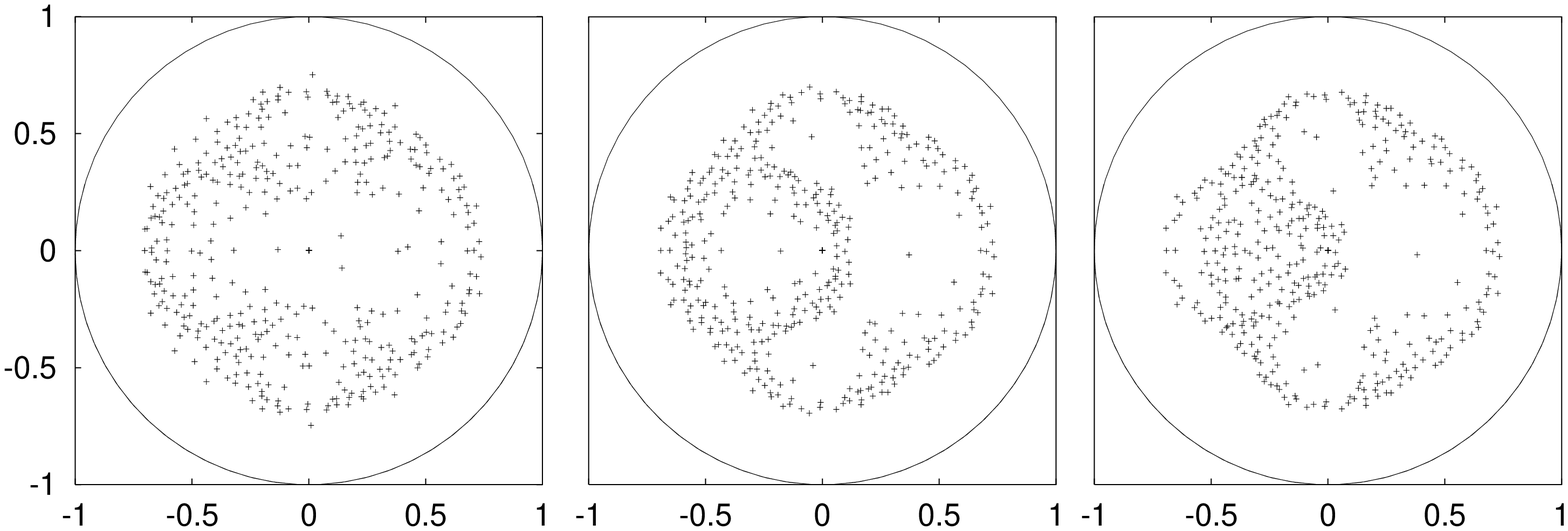}
    \caption{
      Spectra of the convergence matrix for multigrid relaxation
      in a sample gauge field.
      Optimized approximations of order 1, 2, and 3 where used from left to
      right for the Schur complement.
      The top row shows the convergence of the multigrid cycle if a
      full inversion is performed on the coarse grid, and one relaxation 
      step on the fine grid. The bottom row shows the same spectrum if 
      only a single relaxation step is performed on each grid. The
      convergence rate is given by the radius of the smallest
      disk containing eigenvalues. 
      \label{fig:7}}
  \end{center}
  \vspace*{6cm}  
\end{figure}

\subsection{Multigrid relaxation}

Relaxation schemes for solving eq.~(\ref{eq:2}) make use of an
iterative process with an update step that is derived from the residual
\begin{equation}
  \label{eq:12}
  r^{(n)} = a - M f^{(n)} \quad,
\end{equation}
where $f^{(n)}$ is the current approximate solution.
The true error, i.e.~the amount by which $f^{(n)}$ has to be updated,
can be calculated from the residual by
\begin{equation}
  \label{eq:13}
  e^{(n)} = f - f^{(n)} = M^{-1} r^{(n)} \quad.
\end{equation}
Given an approximation $\bar M^{-1}$ to $M^{-1}$, 
an approximate update step reads
\begin{equation}
  \label{eq:13a}
  f^{(n+1)} = f^{(n)} + T r^{(n)} = (1 - \bar M^{-1} M) f^{(n)} + T a \quad.
\end{equation}
This iteration always has a fixed point at the true solution $f$, and
the quality of the approximation $\bar M^{-1}$ to $M^{-1}$ only
determines the convergence characteristics of the process.

Multigrid relaxation is based on using the Schur complement along with
restriction and interpolation to construct an approximation to
$M^{-1}$ on the coarse lattice.  First, the residual is restricted to
the coarse lattice using the restriction
\begin{equation}
  \label{eq:62}
  \bar r^{(n)} = e_1 + Q^\conj e_2 \quad.
\end{equation}
A coarse lattice approximation $\bar e^{(n)}$ to the true error is
then found by applying the inverse of the approximate Schur
complement:
\begin{equation}
  \label{eq:63}
  \bar e^{(n)} = \bar S^{-1} \bar r^{(n)} \quad.
\end{equation}
Finally, this is interpolated back to the fine lattice:
\begin{equation}
  \label{eq:64}
  \bar e_1^{(n)} = \bar e^{(n)} \quad, \qquad
  \bar e_2^{(n)} = Q \bar e^{(n)}
\end{equation}
and used to update the approximation.  This update step acts on the
coarse degrees of freedom only and must be followed by a compatible
relaxation step on the fine lattice.  In matrix form, the
approximate inverse of $M$ used here reads
\begin{equation}
  \label{eq:66}
  T = 
  \left(
    \begin{array}{cc}
      1 & 0 \\ \bar Q & 1
    \end{array}
  \right) \, \left(
    \begin{array}{cc}
      \bar S^{-1} & 0 \\ 0 & 0
    \end{array}
  \right) \, \left(
    \begin{array}{cc}
      1 & \bar Q \\  0 & 1
    \end{array}
  \right) \quad.
\end{equation}
It differs from the true LU decomposition (\ref{eq:44}) in that the
lower right entry of the matrix in the middle is zero instead of
$M^{-1}_{22}$.  If a perfect interpolation and restriction is chosen,
$\bar Q = -M_{22}^{-1} M_{21}$, the iteration matrix for the coarse-grid
step reads
\begin{equation}
  \label{eq:67}
  1-TM = \left(
    \begin{array}{cc}
      1 - \bar S^{-1} S & 0 \\ Q \bar S^{-1} S & 1
    \end{array}
    \right) \quad.
\end{equation}
Its convergence properties on the coarse lattice are thus given again
by $1 - \bar S^{-1} S$ as in the case of preconditioning; on the fine
lattice, it performs no relaxation at all, and must be followed by an
ordinary relaxation step using $M_{22}$.

\begin{figure}[htb]
  \begin{center}
    \includegraphics[width=.8\hsize]{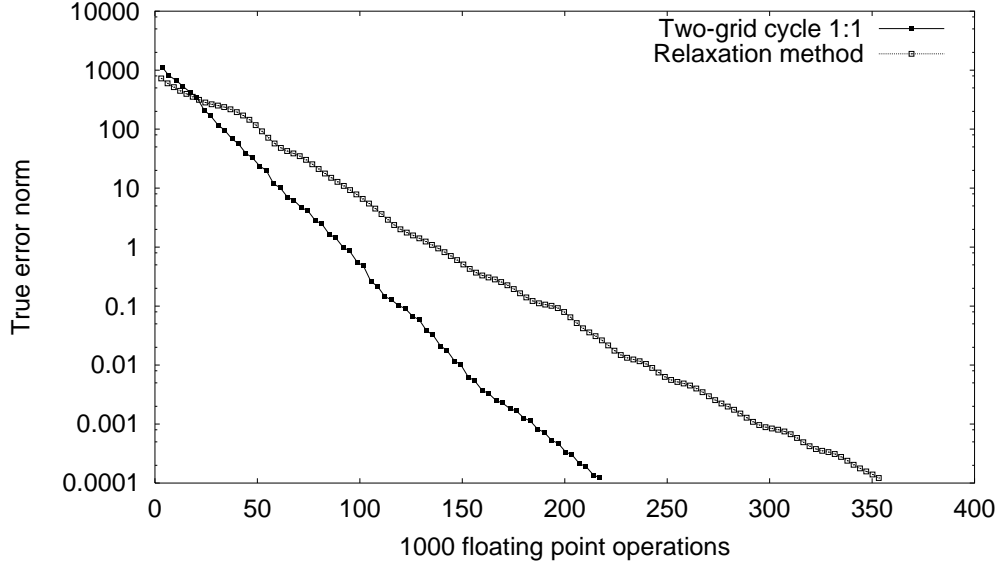}    
    \caption{Sample convergence of a two-grid relaxation method 
      compared to the same
      method applied to the base grid, for a $U(1)$ Dirac operator on
      a $16 \times 16$-lattice.  The horizontal axis gives the number
      of floating-point multiplications, the vertical axis the true
      error of the algorithm. For the two-grid algorithm, one sweep on
      the coarse lattice and one sweep on the fine lattice were
      performed alternately.}
    \label{fig:8}
  \end{center}
\end{figure}

Fig.~\ref{fig:7} shows the spectra of the convergence matrix for
multigrid relaxation. In the top row, each step consists of performing
a complete relaxation on the coarse lattice, and one relaxation step
on the fine lattice. This leads to a concentration of eigenvalues
around zero, showing modes that are nearly completely reduced on the
coarse grid. The remaining modes must be reduced on the fine grid and
determine the residual convergence rate.  In an actual application,
one would not completely reduce the modes on the coarse grid, but
perform coarse- and fine-grid relaxation alternately. The resulting
convergence spectrum is shown in the bottom row; it has a similar
convergence rate as the complete reduction, but requires much less
operations. Note that increasing the order of the approximation from 2
to 3 does not improve the convergence, but this might be related to
the fact that only one fine-grid relaxation step was used here.

Finally, we compare the number of operations required to calculate a
sample Green's function in $U(1)$ lattice gauge theory on a $16\times
16$ lattice. It must be cautioned that this is only an example
calculation using two-grid relaxation, not a full multigrid algorithm.
The method offers a multitude of choices and tunable parameters: the
order and type of the approximation for the $LU$ decomposition, the
iterative algorithm (used here is for simplicity Jacobi relaxation; in
actual applications one would probable use Gauss-Seidel), the
multigrid cycle, and the convergence criteria on the two grids. The
size and dimensionality of the lattice and the gauge group, as well as
the amount of disorder in the gauge field, have been seen to
decisively influence results in other algorithms. Also, the two-grid
algorithm must be extended to a true multigrid algorithm by
recursively applying the method. And finally relaxation might be less
performant than a preconditioned Krylov solver. The ultimate measure
of performance will be the actual computer time used in the
algorithms, but this will depend on the implementation and on the
architecture of the target machine. For example, the computation of
the approximate Schur complement can either be performed beforehand
and stored in memory, or on the fly in the matrix-vector
multiplication of the Schur complement, depending on the amount of
memory available. Here a lower order of approximation, resulting in a
smaller Schur complement, can be favorable though its convergence rate
per relaxation step might be slower.

With these precautions, an example run for calculating a sample
Green's function is shown in Fig.~\ref{fig:8}. As a measure of
computer time used, the approximate number of floating-point
multiplications is given. The convergence is measured as the
error norm against the true result.

\section{Conclusions}

It was demonstrated how the algebraic multigrid method 
can be applied to disordered linear operators.
Other than in previous multigrid approaches, a thinned lattice for the
coarse degrees of freedom is used and averaging over block spins is
avoided. The interpolation operator is chosen to obtain a block $LU$
decomposition, and the resulting coarse-grid dynamics is completely
described by the Schur complement of the original matrix. Gauge
covariance is automatically taken into account by this procedure.

The Schur complement can be approximated to arbitrary precision in a
numerical optimization process by expanding it in a linear basis
constructed from connected paths on the original lattice, forming a
generalized stencil for disordered operators.  Similar to
renormalization group techniques, the numerical procedure for
obtaining the expansion coefficients uses the coarse-grid projection
of an ensemble of systems to obtain the coefficients of the
approximate coarse-grid operator.  In this way, the information
gathered in the optimization process can be used to speed up the
actual calculations as the optimized effective coarse-grid operator
``learns'' the dynamics of the system.

The resulting effective coarse-grid operator is constructed from the
generalized stencil as a weighted sum over paths on the original
lattice. It forms a denser, but smaller matrix with next-neighbor and
higher interactions and can be used to improve performance in
numerical algorithms by preconditioning or similar methods, hopefully
not only for matrix inversion, but also for other problems such as
eigensystem analysis and rational matrix functions.  A variety of
tunable parameters make it possible to adapt the procedure to
different algorithms and architectures.  Whether the method will
actually improve real-life algorithms, remains of course to be seen
until realistic systems and efficient implementations are
investigated.

\noindent{\small{\bf Acknowledgements}\\
  The author wishes to thank K.~Schilling, T.~Lippert, B.~Medeke,
  and J.~Negele for useful discussions, and the
  Center for Theoretical Physics at MIT for its hospitality. Computer
  time was provided by the NICse cluster computer at the John von
  Neumann Institute.}

\ifcbtex
\raggedright
\bibliographystyle{cb}
\fi
\bibliography{mg}

\begin{thebibliography}{10}
\expandafter\ifx\csname bibnamefont\endcsname\relax
  \def\bibnamefont#1{#1}\fi
\expandafter\ifx\csname bibfnamefont\endcsname\relax
  \def\bibfnamefont#1{#1}\fi
\expandafter\ifx\csname url\endcsname\relax
  \def\url#1{\texttt{#1}}\fi
\expandafter\ifx\csname urlprefix\endcsname\relax\def\urlprefix{URL }\fi
\providecommand{\bibinfo}[2]{#2}
\providecommand{\eprint}[2][]{\url{#2}}

\bibitem{heplat-9404013}
\bibinfo{author}{\bibfnamefont{A.}~\bibnamefont{Frommer}},
  \bibinfo{author}{\bibfnamefont{V.}~\bibnamefont{Hannemann}},
  \bibinfo{author}{\bibfnamefont{B.}~\bibnamefont{N{\"o}ckel}},
  \bibinfo{author}{\bibfnamefont{T.}~\bibnamefont{Lippert}}, \bibnamefont{and}
  \bibinfo{author}{\bibfnamefont{K.}~\bibnamefont{Schilling}},
  \emph{\bibinfo{title}{Accelerating the wilson fermion matrix inversions by
  means of the stabilized biconjugate gradient algorithm}},
  \bibinfo{journal}{Intl. J. Mod. Phys. C} \textbf{\bibinfo{volume}{5}},
  \bibinfo{pages}{1073} (\bibinfo{year}{1994}), \eprint{hep-lat/9404013}.

\bibitem{heplat-9608074}
\bibinfo{author}{\bibfnamefont{A.}~\bibnamefont{Frommer}},
  \emph{\bibinfo{title}{Linear systems solvers - recent developments and
  implications for lattice computations}}, \bibinfo{journal}{Nucl. Phys. (Proc.
  Suppl.)} \textbf{\bibinfo{volume}{53}}, \bibinfo{pages}{120}
  (\bibinfo{year}{1997}), \eprint{hep-lat/9608074}.

\bibitem{gutknecht-wup-1999}
\bibinfo{author}{\bibfnamefont{M.~H.} \bibnamefont{Gutknecht}},
  \emph{\bibinfo{title}{On {Lanczos}-type methods for {Wilson} fermions}}, in
  \bibinfo{editor}{\bibnamefont{\bibinfo{editor}{\bibnamefont{\bibinfo{editor}%
{Frommer}}} \emph{et~al.}}}  \cite{wuppertal-1999-proceedings}.

\bibitem{heplat-9602019}
\bibinfo{author}{\bibfnamefont{S.}~\bibnamefont{Fischer}},
  \bibinfo{author}{\bibfnamefont{A.}~\bibnamefont{Frommer}},
  \bibinfo{author}{\bibfnamefont{U.}~\bibnamefont{Gl{\"a}ssner}},
  \bibinfo{author}{\bibfnamefont{T.}~\bibnamefont{Lippert}},
  \bibinfo{author}{\bibfnamefont{G.}~\bibnamefont{Ritzenh{\"o}fer}},
  \bibnamefont{and}
  \bibinfo{author}{\bibfnamefont{K.}~\bibnamefont{Schilling}},
  \emph{\bibinfo{title}{A parallel {SSOR} preconditioner for lattice {QCD}}},
  \bibinfo{journal}{Comp. Phys. Comm.} \textbf{\bibinfo{volume}{98}},
  \bibinfo{pages}{20} (\bibinfo{year}{1998}), \eprint{hep-lat/9602019}.

\bibitem{heplat-0011080}
\bibinfo{author}{\bibfnamefont{M.}~\bibnamefont{Peardon}},
  \emph{\bibinfo{title}{Accelerating the {Hybrid Monte Carlo} algorithm with
  {ILU} preconditioning}}  (\bibinfo{year}{2000}), \eprint{hep-lat/0011080}.

\bibitem{hackbusch-springer}
\bibinfo{author}{\bibfnamefont{W.}~\bibnamefont{Hackbusch}},
  \emph{\bibinfo{title}{Multi-Grid Methods and Applications}},
  no.~\bibinfo{number}{4} in \bibinfo{series}{Springer Series in Computational
  Mathematics} (\bibinfo{publisher}{Springer}, \bibinfo{address}{Berlin},
  \bibinfo{year}{1985}).

\bibitem{mccormick-siam}
\bibinfo{editor}{\bibfnamefont{S.~F.} \bibnamefont{McCormick}}, ed.,
  \emph{\bibinfo{title}{Multigrid Methods}} (\bibinfo{publisher}{SIAM,
  Philadelphia}, \bibinfo{year}{1987}).

\bibitem{mack-1988}
\bibinfo{author}{\bibfnamefont{G.}~\bibnamefont{Mack}},
  \emph{\bibinfo{title}{Multigrid methods in quantum field theory}}, in
  \emph{\bibinfo{booktitle}{Nonperturbative Quantum Field Theory}}, edited by
  \bibinfo{editor}{\bibnamefont{{G. 't Hooft et al.}}}
  (\bibinfo{publisher}{Plenum Press, New York}, \bibinfo{year}{1988}).

\bibitem{juelich-1991}
\bibinfo{editor}{\bibfnamefont{H.~J.} \bibnamefont{Herrmann}} \bibnamefont{and}
  \bibinfo{editor}{\bibfnamefont{F.}~\bibnamefont{Karsch}}, eds.,
  \emph{\bibinfo{title}{Workshop on Fermion Algorithms}}
  (\bibinfo{publisher}{World Scientific, Singapore}, \bibinfo{year}{1991}).

\bibitem{prd-43-1965}
\bibinfo{author}{\bibfnamefont{R.~C.} \bibnamefont{Brower}},
  \bibinfo{author}{\bibfnamefont{C.}~\bibnamefont{Rebbi}}, \bibnamefont{and}
  \bibinfo{author}{\bibfnamefont{E.}~\bibnamefont{Vicari}},
  \emph{\bibinfo{title}{Projective multigrid for propagators in lattice gauge
  theory}}, \bibinfo{journal}{Phys. Rev. D} \textbf{\bibinfo{volume}{43}},
  \bibinfo{pages}{1965} (\bibinfo{year}{1991}).

\bibitem{brower-moriarty-rebbi-vicari}
\bibinfo{author}{\bibfnamefont{R.~C.} \bibnamefont{Brower}},
  \bibinfo{author}{\bibfnamefont{K.~J.~M.} \bibnamefont{Moriarty}},
  \bibinfo{author}{\bibfnamefont{C.}~\bibnamefont{Rebbi}}, \bibnamefont{and}
  \bibinfo{author}{\bibfnamefont{E.}~\bibnamefont{Vicari}},
  \emph{\bibinfo{title}{Multigrid propagators in the presence of disordered
  {$U(1)$} gauge fields}}, \bibinfo{journal}{Phys. Rev. D}
  \textbf{\bibinfo{volume}{43}}(\bibinfo{number}{6}), \bibinfo{pages}{1974}
  (\bibinfo{year}{1991}).

\bibitem{brower-edwards-rebbi-1991}
\bibinfo{author}{\bibfnamefont{R.~C.} \bibnamefont{Brower}},
  \bibinfo{author}{\bibfnamefont{R.~G.} \bibnamefont{Edwards}},
  \bibinfo{author}{\bibfnamefont{C.}~\bibnamefont{Rebbi}}, \bibnamefont{and}
  \bibinfo{author}{\bibfnamefont{E.}~\bibnamefont{Vicari}},
  \emph{\bibinfo{title}{Projective multigrid for wilson fermions}},
  \bibinfo{journal}{Nucl. Phys. B} \textbf{\bibinfo{volume}{366}},
  \bibinfo{pages}{689} (\bibinfo{year}{1991}).

\bibitem{vyas-jue-1991}
\bibinfo{author}{\bibfnamefont{V.}~\bibnamefont{Vyas}},
  \emph{\bibinfo{title}{Calculating the quark propagator using the
  {Migdal-Kadanoff} transformation}}, in
  \bibinfo{editor}{\bibnamefont{\bibinfo{editor}{Herrmann} and
  \bibinfo{editor}{Karsch}}}  \cite{juelich-1991}, pp.
  \bibinfo{pages}{169--172}.

\bibitem{prd-vyas}
\bibinfo{author}{\bibfnamefont{V.}~\bibnamefont{Vyas}},
  \emph{\bibinfo{title}{Real-space renormalization-group approach to the
  multigrid method}}, \bibinfo{journal}{Phys. Rev. D}
  \textbf{\bibinfo{volume}{43}}(\bibinfo{number}{10}), \bibinfo{pages}{3465}
  (\bibinfo{year}{1991}).

\bibitem{plb-253-185}
\bibinfo{author}{\bibfnamefont{R.}~\bibnamefont{Ben-Av}},
  \bibinfo{author}{\bibfnamefont{A.}~\bibnamefont{Brandt}},
  \bibinfo{author}{\bibfnamefont{M.}~\bibnamefont{Harmatz}},
  \bibinfo{author}{\bibfnamefont{E.}~\bibnamefont{Katznelson}},
  \bibinfo{author}{\bibfnamefont{P.~G.} \bibnamefont{Lauwers}},
  \bibinfo{author}{\bibfnamefont{S.}~\bibnamefont{Solomon}}, \bibnamefont{and}
  \bibinfo{author}{\bibfnamefont{K.}~\bibnamefont{Wolowesky}},
  \emph{\bibinfo{title}{Fermion simulations using parallel transported
  multigrid}}, \bibinfo{journal}{Phys. Lett. B} \textbf{\bibinfo{volume}{253}},
  \bibinfo{pages}{185} (\bibinfo{year}{1991}).

\bibitem{solomon-jue-1991}
\bibinfo{author}{\bibfnamefont{S.}~\bibnamefont{Solomon}} \bibnamefont{and}
  \bibinfo{author}{\bibfnamefont{P.~G.} \bibnamefont{Lauwers}},
  \emph{\bibinfo{title}{Parallel-transported multigrid beats conjugate
  gradient}}, in \bibinfo{editor}{\bibnamefont{\bibinfo{editor}{Herrmann} and
  \bibinfo{editor}{Karsch}}}  \cite{juelich-1991}, pp.
  \bibinfo{pages}{149--160}.

\bibitem{heplat-9204014}
\bibinfo{author}{\bibfnamefont{A.}~\bibnamefont{Brandt}},
  \emph{\bibinfo{title}{Multigrid methods in lattice field computations}},
  \bibinfo{journal}{Nucl. Phys. B (Proc. Suppl.)}
  \textbf{\bibinfo{volume}{26}}, \bibinfo{pages}{137} (\bibinfo{year}{1992}),
  \eprint{hep-lat/9204014}.

\bibitem{lauwers-benav-solomon-1992}
\bibinfo{author}{\bibfnamefont{P.~G.} \bibnamefont{Lauwers}},
  \bibinfo{author}{\bibfnamefont{R.}~\bibnamefont{Ben-Av}}, \bibnamefont{and}
  \bibinfo{author}{\bibfnamefont{S.}~\bibnamefont{Solomon}},
  \emph{\bibinfo{title}{Inverting the dirac matrix for {$SU(2)$} lattice gauge
  theory by means of parallel transported multigrid}}, \bibinfo{journal}{Nucl.
  Phys. B} \textbf{\bibinfo{volume}{374}}, \bibinfo{pages}{249}
  (\bibinfo{year}{1992}).

\bibitem{hulsebos-jue-1991}
\bibinfo{author}{\bibfnamefont{J.~S.} \bibnamefont{A.~Hulsebos}}
  \bibnamefont{and} \bibinfo{author}{\bibfnamefont{J.~C.} \bibnamefont{Vink}},
  \emph{\bibinfo{title}{Multigrid inversion of the staggered fermion matrix
  with {$U(1)$} and {$SU(2)$} gauge fields}}, in
  \bibinfo{editor}{\bibnamefont{\bibinfo{editor}{Herrmann} and
  \bibinfo{editor}{Karsch}}}  \cite{juelich-1991}, pp.
  \bibinfo{pages}{161--168}.

\bibitem{npb-331-531}
\bibinfo{author}{\bibfnamefont{A.}~\bibnamefont{Hulsebos}},
  \bibinfo{author}{\bibfnamefont{J.}~\bibnamefont{Smit}}, \bibnamefont{and}
  \bibinfo{author}{\bibfnamefont{J.~C.} \bibnamefont{Vink}},
  \emph{\bibinfo{title}{Multigrid {Monte Carlo} for a {Bose} field in an
  external gauge field}}, \bibinfo{journal}{Nucl. Phys. B}
  \textbf{\bibinfo{volume}{331}}, \bibinfo{pages}{531} (\bibinfo{year}{1990}).

\bibitem{plb-272-81}
\bibinfo{author}{\bibfnamefont{J.~C.} \bibnamefont{Vink}},
  \emph{\bibinfo{title}{Multigrid inversion of staggered and wilson fermion
  operators with {$SU(2)$} gauge fields in two dimensions}},
  \bibinfo{journal}{Phys. Lett. B} \textbf{\bibinfo{volume}{272}},
  \bibinfo{pages}{81} (\bibinfo{year}{1991}).

\bibitem{amsterdam-1991}
\bibinfo{author}{\bibfnamefont{A.}~\bibnamefont{Hulsebos}},
  \bibinfo{author}{\bibfnamefont{J.}~\bibnamefont{Smit}}, \bibnamefont{and}
  \bibinfo{author}{\bibfnamefont{J.~C.} \bibnamefont{Vink}},
  \emph{\bibinfo{title}{Multigrid inversion of lattice fermion operators}},
  \bibinfo{journal}{Nucl. Phys. B} \textbf{\bibinfo{volume}{368}},
  \bibinfo{pages}{379} (\bibinfo{year}{1992}).

\bibitem{kalkreuter-jue-1991}
\bibinfo{author}{\bibfnamefont{T.}~\bibnamefont{Kalkreuter}},
  \emph{\bibinfo{title}{Blockspin and multigrid for staggered fermions in
  non-{Abelian} gauge fields}}, in
  \bibinfo{editor}{\bibnamefont{\bibinfo{editor}{Herrmann} and
  \bibinfo{editor}{Karsch}}}  \cite{juelich-1991}, pp.
  \bibinfo{pages}{121--148}.

\bibitem{heplat-9304004}
\bibinfo{author}{\bibfnamefont{T.}~\bibnamefont{Kalkreuter}},
  \emph{\bibinfo{title}{Idealized multigrid algorithm for staggered fermions}},
  \bibinfo{journal}{Phys. Rev. D} \textbf{\bibinfo{volume}{48}},
  \bibinfo{pages}{1926} (\bibinfo{year}{1993}), \eprint{hep-lat/9304004}.

\bibitem{heplat-9310029}
\bibinfo{author}{\bibfnamefont{T.}~\bibnamefont{Kalkreuter}},
  \emph{\bibinfo{title}{Towards multigrid methods for propagators of staggered
  fermions with improved averaging and interpolation operators}},
  \bibinfo{journal}{Nucl. Phys. B (Proc. Suppl.)}
  \textbf{\bibinfo{volume}{34}}, \bibinfo{pages}{768} (\bibinfo{year}{1994}),
  \eprint{hep-lat/9310029}.

\bibitem{heplat-9409008}
\bibinfo{author}{\bibfnamefont{T.}~\bibnamefont{Kalkreuter}},
  \emph{\bibinfo{title}{Multigrid methods for lattice gauge theories}},
  \bibinfo{journal}{J. Comp. Appl. Math.} \textbf{\bibinfo{volume}{63}},
  \bibinfo{pages}{57} (\bibinfo{year}{1995}), \eprint{hep-lat/9409008}.

\bibitem{heplat-9408013}
\bibinfo{author}{\bibfnamefont{T.}~\bibnamefont{Kalkreuter}},
  \emph{\bibinfo{title}{Spectrum of the {Dirac} operator and multigrid
  algorithm with dynamical staggered fermions}}, \bibinfo{journal}{Phys. Rev.
  D} \textbf{\bibinfo{volume}{51}}, \bibinfo{pages}{1305}
  (\bibinfo{year}{1995}), \eprint{hep-lat/9408013}.

\bibitem{thesis-baeker}
\bibinfo{author}{\bibfnamefont{M.}~\bibnamefont{B{\"a}ker}},
  \emph{\bibinfo{title}{A multiscale view of propagators in gauge fields}},
  Ph.D. thesis, \bibinfo{school}{Universit{\"a}t Hamburg}
  (\bibinfo{year}{1995}).

\bibitem{wuppertal-1999-proceedings}
\bibinfo{editor}{\bibfnamefont{A.}~\bibnamefont{Frommer}},
  \bibinfo{editor}{\bibfnamefont{T.}~\bibnamefont{Lippert}},
  \bibinfo{editor}{\bibfnamefont{B.}~\bibnamefont{Medeke}}, \bibnamefont{and}
  \bibinfo{editor}{\bibfnamefont{K.}~\bibnamefont{Schilling}}, eds.,
  \emph{\bibinfo{title}{Numerical Challenges in Lattice Quantum
  Chromodynamics}} (\bibinfo{publisher}{Springer, Berlin},
  \bibinfo{year}{2000}).

\bibitem{siam-ruge-stueben}
\bibinfo{author}{\bibfnamefont{J.~W.} \bibnamefont{Ruge}} \bibnamefont{and}
  \bibinfo{author}{\bibfnamefont{K.}~\bibnamefont{St{\"u}ben}},
  \emph{\bibinfo{title}{Algebraic multigrid}}, in
  \bibinfo{editor}{\bibnamefont{\bibinfo{editor}{McCormick}}}
  \cite{mccormick-siam}, p.~\bibinfo{pages}{73}.

\bibitem{reusken-wup-1999}
\bibinfo{author}{\bibfnamefont{A.}~\bibnamefont{Reusken}},
  \emph{\bibinfo{title}{An algebraic multilevel preconditioner for symmetric
  positive definite and indefinite problems}}, in
  \bibinfo{editor}{\bibnamefont{\bibinfo{editor}{\bibnamefont{\bibinfo{editor}%
{Frommer}}} \emph{et~al.}}}  \cite{wuppertal-1999-proceedings}, pp.
  \bibinfo{pages}{66--83}.

\bibitem{notay-wup-1999}
\bibinfo{author}{\bibfnamefont{Y.}~\bibnamefont{Notay}},
  \emph{\bibinfo{title}{On algebraic multilevel preconditioning}}, in
  \bibinfo{editor}{\bibnamefont{\bibinfo{editor}{\bibnamefont{\bibinfo{editor}%
{Frommer}}} \emph{et~al.}}}  \cite{wuppertal-1999-proceedings}, pp.
  \bibinfo{pages}{84--98}.

\bibitem{brandt-2000}
\bibinfo{author}{\bibfnamefont{A.}~\bibnamefont{Brandt}},
  \emph{\bibinfo{title}{General highly accurate algebraic coarsening}},
  \bibinfo{journal}{Electronic Transactions on Numerical Analysis}
  \textbf{\bibinfo{volume}{10}}, \bibinfo{pages}{1} (\bibinfo{year}{2000}).

\bibitem{zhang-2000}
\bibinfo{author}{\bibfnamefont{J.}~\bibnamefont{Zhang}},
  \emph{\bibinfo{title}{On preconditioning {Schur} complement and {Schur}
  complement preconditioning}}, \bibinfo{journal}{Electronic Transactions on
  Numerical Analysis} \textbf{\bibinfo{volume}{10}}, \bibinfo{pages}{115}
  (\bibinfo{year}{2000}).

\bibitem{prl-61-1333}
\bibinfo{author}{\bibfnamefont{R.~G.} \bibnamefont{Edwards}},
  \bibinfo{author}{\bibfnamefont{J.}~\bibnamefont{Goodman}}, \bibnamefont{and}
  \bibinfo{author}{\bibfnamefont{A.~D.} \bibnamefont{Sokal}},
  \emph{\bibinfo{title}{Multigrid method for the random-resistor problem}},
  \bibinfo{journal}{Phys. Rev. Lett.}
  \textbf{\bibinfo{volume}{61}}(\bibinfo{number}{12}), \bibinfo{pages}{1333}
  (\bibinfo{year}{1988}).

\bibitem{medeke-wup-1999}
\bibinfo{author}{\bibfnamefont{B.}~\bibnamefont{Medeke}},
  \emph{\bibinfo{title}{On algebraic multilevel preconditioners in lattice
  gauge theory}}, in
  \bibinfo{editor}{\bibnamefont{\bibinfo{editor}{\bibnamefont{\bibinfo{editor}%
{Frommer}}} \emph{et~al.}}}  \cite{wuppertal-1999-proceedings}, pp.
  \bibinfo{pages}{99--114}.

\bibitem{condmat-0009449}
\bibinfo{author}{\bibfnamefont{Q.}~\bibnamefont{Hou}},
  \bibinfo{author}{\bibfnamefont{N.}~\bibnamefont{Goldenfeld}},
  \bibnamefont{and} \bibinfo{author}{\bibfnamefont{A.}~\bibnamefont{McKane}},
  \emph{\bibinfo{title}{Renormalization group and perfect operators for
  stochastic differential equations}}  (\bibinfo{year}{2000}),
  \eprint{cond-mat/0009449}.

\bibitem{kuti-neumann-series}
\bibinfo{author}{\bibfnamefont{J.}~\bibnamefont{Kuti}},
  \emph{\bibinfo{title}{Stochastic method for the numerical study of lattice
  fermions}}, \bibinfo{journal}{Phys. Rev. Lett.}
  \textbf{\bibinfo{volume}{49}}(\bibinfo{number}{3}), \bibinfo{pages}{183}
  (\bibinfo{year}{1982}).

\bibitem{vyas-random-walk}
\bibinfo{author}{\bibfnamefont{V.}~\bibnamefont{Vyas}},
  \emph{\bibinfo{title}{Random walk representation of the lattice fermionic
  propagators and the quark model}}  (\bibinfo{year}{1997}),
  \eprint{hep-lat/9101010}.

\bibitem{kehr-1996}
\bibinfo{author}{\bibfnamefont{K.~W.} \bibnamefont{Kehr}} \bibnamefont{and}
  \bibinfo{author}{\bibfnamefont{T.}~\bibnamefont{Wichmann}},
  \emph{\bibinfo{title}{Diffusion coefficients of single and many particles in
  lattices with different forms of disorder}}, \bibinfo{journal}{Materials
  Science Forum} \textbf{\bibinfo{volume}{223--4}}, \bibinfo{pages}{151}
  (\bibinfo{year}{1996}).

\bibitem{guo-miller-2000}
\bibinfo{author}{\bibfnamefont{H.}~\bibnamefont{Guo}} \bibnamefont{and}
  \bibinfo{author}{\bibfnamefont{B.~N.} \bibnamefont{Miller}},
  \emph{\bibinfo{title}{{Monte Carlo} study of localization on a
  one-dimensional lattice}}, \bibinfo{journal}{J. Stat. Phys}
  \textbf{\bibinfo{volume}{98}}, \bibinfo{pages}{347} (\bibinfo{year}{2000}).

\bibitem{moulton-knapek-dendy-1998}
\bibinfo{author}{\bibfnamefont{J.~D.} \bibnamefont{Moulton}},
  \bibinfo{author}{\bibfnamefont{S.}~\bibnamefont{Knapek}}, \bibnamefont{and}
  \bibinfo{author}{\bibfnamefont{J.~E.} \bibnamefont{Dendy}},
  \emph{\bibinfo{title}{Multilevel upscaling in heterogeneous porous media}},
  \bibinfo{type}{Tech. Rep.}, \bibinfo{institution}{Center for Nonlinear
  Science, Los Alamos National Laboratory} (\bibinfo{year}{1999}).

\bibitem{SKnapek:99b}
\bibinfo{author}{\bibfnamefont{S.}~\bibnamefont{Knapek}},
  \emph{\bibinfo{title}{Matrix-dependent multigrid homogenization for diffusion
  problems}}, \bibinfo{journal}{SIAM J.~Sci.~Comp.}
  \textbf{\bibinfo{volume}{20}}(\bibinfo{number}{2}), \bibinfo{pages}{515}
  (\bibinfo{year}{1999}).

\bibitem{iwcc-melbourne}
\bibinfo{editor}{\bibfnamefont{R.}~\bibnamefont{Buyya}},
  \bibinfo{editor}{\bibfnamefont{M.}~\bibnamefont{Baker}},
  \bibinfo{editor}{\bibfnamefont{K.}~\bibnamefont{Hawick}}, \bibnamefont{and}
  \bibinfo{editor}{\bibfnamefont{H.}~\bibnamefont{James}}, eds.,
  \emph{\bibinfo{title}{Proceedings of the 1st International Workshop on
  Cluster Computing}} (\bibinfo{publisher}{IEEE Computer Society},
  \bibinfo{address}{Melbourne, Australia}, \bibinfo{year}{1999}).

\bibitem{iwcc-melbourne-cbest}
\bibinfo{author}{\bibfnamefont{C.}~\bibnamefont{Best}},
  \bibinfo{author}{\bibfnamefont{N.}~\bibnamefont{Eicker}},
  \bibinfo{author}{\bibfnamefont{T.}~\bibnamefont{Lippert}},
  \bibinfo{author}{\bibfnamefont{M.}~\bibnamefont{Peardon}},
  \bibinfo{author}{\bibfnamefont{P.}~\bibnamefont{{\"U}berholz}},
  \bibnamefont{and}
  \bibinfo{author}{\bibfnamefont{K.}~\bibnamefont{Schilling}},
  \emph{\bibinfo{title}{Lattice field theory on cluster computers: Vector- vs.
  cache-centric programming}}, in
  \bibinfo{editor}{\bibnamefont{\bibinfo{editor}{\bibnamefont{\bibinfo{editor}%
{Buyya}}} \emph{et~al.}}}  \cite{iwcc-melbourne}.

\bibitem{cs-0007027}
\bibinfo{author}{\bibfnamefont{M.~A.} \bibnamefont{Frumkin}} \bibnamefont{and}
  \bibinfo{author}{\bibfnamefont{R.~F.~V.} \bibnamefont{der Wijngaart}},
  \emph{\bibinfo{title}{Efficient cache use for stencil operations on
  structured discretization grids}}  (\bibinfo{year}{2000}),
  \eprint{cs.PF/0007027}.

\bibitem{heplat-9709130}
\bibinfo{author}{\bibfnamefont{T.}~\bibnamefont{Ivanenko}} \bibnamefont{and}
  \bibinfo{author}{\bibfnamefont{J.}~\bibnamefont{Negele}},
  \emph{\bibinfo{title}{Evidence of instanton effects in hadrons from the study
  of low eigenfunctions of the {Dirac} operator}}, \bibinfo{journal}{Nucl.
  Phys. B (Proc. Suppl.)} \textbf{\bibinfo{volume}{63}}, \bibinfo{pages}{504}
  (\bibinfo{year}{1998}), \eprint{hep-lat/9709130}.

\bibitem{heplat-0010049}
\bibinfo{author}{\bibfnamefont{H.}~\bibnamefont{M.~G{\"o}ckeler}},
  \bibinfo{author}{\bibfnamefont{P.}~\bibnamefont{Rakow}},
  \bibinfo{author}{\bibfnamefont{A.}~\bibnamefont{Sch{\"a}fer}},
  \bibinfo{author}{\bibfnamefont{W.}~\bibnamefont{S{\"o}ldner}},
  \bibnamefont{and} \bibinfo{author}{\bibfnamefont{T.}~\bibnamefont{Wettig}},
  \emph{\bibinfo{title}{Dirac eigenvalues and eigenvectors at finite
  temperature}}  (\bibinfo{year}{2000}), \eprint{hep-lat/0010049}.

\bibitem{heplat-0003021}
\bibinfo{author}{\bibfnamefont{P.~H.} \bibnamefont{Damgaard}},
  \bibinfo{author}{\bibfnamefont{U.~M.} \bibnamefont{Heller}},
  \bibinfo{author}{\bibfnamefont{R.}~\bibnamefont{Niclasen}}, \bibnamefont{and}
  \bibinfo{author}{\bibfnamefont{K.}~\bibnamefont{Rummukainen}},
  \emph{\bibinfo{title}{Low-lying eigenvalues of the {QCD} {Dirac} operator at
  finite temperature}}  (\bibinfo{year}{2000}), \eprint{hep-lat/0003021}.

\bibitem{heplat-9806025}
\bibinfo{author}{\bibfnamefont{H.}~\bibnamefont{Neuberger}},
  \emph{\bibinfo{title}{A practical implementation of the {Overlap-Dirac}
  operator}}, \bibinfo{journal}{Phys. Rev. Lett.}
  \textbf{\bibinfo{volume}{81}}, \bibinfo{pages}{4060} (\bibinfo{year}{1998}).

\bibitem{jansen-wup-1999}
\bibinfo{author}{\bibfnamefont{P.}~\bibnamefont{Hernandez}},
  \bibinfo{author}{\bibfnamefont{K.}~\bibnamefont{Jansen}}, \bibnamefont{and}
  \bibinfo{author}{\bibfnamefont{L.}~\bibnamefont{Lellouch}},
  \emph{\bibinfo{title}{A numerical treatment of {Neuberger's} lattice {Dirac}
  operator}}, in
  \bibinfo{editor}{\bibnamefont{\bibinfo{editor}{\bibnamefont{\bibinfo{editor}%
{Frommer}}} \emph{et~al.}}}  \cite{wuppertal-1999-proceedings}.

\bibitem{heplat-0007003}
\bibinfo{author}{\bibfnamefont{A.}~\bibnamefont{Bori{\c c}i}},
  \emph{\bibinfo{title}{Chiral fermions and multigrid}}
  (\bibinfo{year}{2000}), \eprint{hep-lat/0007003}.

\bibitem{prd-lippert}
\bibinfo{author}{\bibfnamefont{I.}~\bibnamefont{Barbour}},
  \bibinfo{author}{\bibfnamefont{E.}~\bibnamefont{Laermann}},
  \bibinfo{author}{\bibfnamefont{T.}~\bibnamefont{Lippert}}, \bibnamefont{and}
  \bibinfo{author}{\bibfnamefont{K.}~\bibnamefont{Schilling}},
  \emph{\bibinfo{title}{Towards the chiral limit with dynamical blocked
  fermions}}, \bibinfo{journal}{Phys. Rev. D}
  \textbf{\bibinfo{volume}{46}}(\bibinfo{number}{8}), \bibinfo{pages}{3618}
  (\bibinfo{year}{1992}).

\bibitem{heplat-0001001}
\bibinfo{author}{\bibfnamefont{W.}~\bibnamefont{Bietenholz}},
  \emph{\bibinfo{title}{Optimizing chirality and scaling of lattice fermions}}
  (\bibinfo{year}{2000}), \eprint{hep-lat/0001001}.

\bibitem{heplat-0003005}
\bibinfo{author}{\bibfnamefont{C.}~\bibnamefont{Gattringer}},
  \emph{\bibinfo{title}{A new approach to {Ginsparg-Wilson} fermions}}
  (\bibinfo{year}{2000}), \eprint{hep-lat/0003005}.

\bibitem{heplat-0003013}
\bibinfo{author}{\bibfnamefont{P.}~\bibnamefont{Hasenfratz}},
  \bibinfo{author}{\bibfnamefont{S.}~\bibnamefont{Hauswirth}},
  \bibinfo{author}{\bibfnamefont{K.}~\bibnamefont{Holland}},
  \bibinfo{author}{\bibfnamefont{T.}~\bibnamefont{J{\"o}rg}},
  \bibinfo{author}{\bibfnamefont{F.}~\bibnamefont{Niedermayer}},
  \bibnamefont{and} \bibinfo{author}{\bibfnamefont{U.}~\bibnamefont{Wenger}},
  \emph{\bibinfo{title}{The construction of generalized dirac operators on the
  lattice}}  (\bibinfo{year}{2000}), \eprint{hep-lat/0003013}.

\end{thebibliography}

\end{document}